\documentclass[12pt]{article}

\usepackage{graphicx}
\usepackage{amssymb}
\usepackage{amsmath}
\usepackage{amsbsy}

\usepackage{booktabs}
\usepackage{siunitx}
\usepackage{caption}
\usepackage{color}
\usepackage[dvipsnames]{xcolor}

\DeclareSIUnit{\calorie}{cal}

\newcommand{\tensor}[1]{\bar{\bar{#1}}}
\newcommand{\unitary}[1]{\boldsymbol{\mathrm{\hat #1}}}
\newcommand{\vecb}[1]{\boldsymbol{\mathrm{#1}}}
\newcommand{\di}{\mathrm{d}} 
\newcommand{\dpart}[2]{\frac{\partial #1}{\partial #2}}

\usepackage{times}

\newenvironment{sciabstract}{%
	\begin{quote} \bf}
	{\end{quote}}

\title{Principles of Hodgkin-Huxley iontronics with two-dimensional nanofluidic memristors}
\title{Memristor effect and neuromorphic behavior in two-dimensional ion transport}
\title{Modeling of emergent memory and voltage spiking in two-dimensional nanofluidics}
\title{Modeling of emergent memory and voltage spiking in two-dimensional iontronics}
\title{Modeling of emergent memory and voltage spiking in two-dimensional ionic transport}
\title{Modeling of emergent memory and voltage spiking in ionic transport through angstr\"om-scale slits}

\author
{Paul Robin$^1$, Nikita Kavokine$^1$, and Lyd\'eric Bocquet$^{1,\ast}$\\
	\\
	\normalsize{$^{1}$Laboratoire de Physique de l'\'Ecole Normale Sup\'erieure,}\\
	\normalsize{ENS, Universit\'e PSL, CNRS, Sorbonne Universit\'e, Universit\'e de Paris, F-75005 Paris, France}\\
	\normalsize{$^\ast$To whom correspondence should be addressed; E-mail:  lyderic.bocquet@ens.fr.}
}

\date{}

\newcommand{\NK}[1]{{\color{black} {#1}}}
\newcommand{\rev}[1]{{\color{black} {#1}}}
\newcommand{\LB}[1]{{\color{black} {#1}}}
\newcommand{\PR}[1]{{\color{black} {#1}}}

\begin{document} 
	
	
	
	
	\maketitle

	
	\begin{sciabstract}
		
		\rev{
			Recent advances in nanofluidics have enabled the confinement of water down to a single molecular layer. Such 
			\NK{monolayer electrolytes show promise in achieving bio-inspired functionalities through molecular control of ion transport.}
			However, \NK{the understanding of ion dynamics in these systems is still scarce}.
			\NK{Here}, we develop an analytical theory, backed up by molecular dynamics simulations, predicting \NK{strongly} nonlinear \NK{effects in} ion transport across \LB{quasi-two-dimensional} slits. We show that under an electric field, ions assemble into elongated clusters, whose slow dynamics result in hysteretic \LB{conduction}. 
			This phenomenon, known as memristor effect, can be harnessed to build \NK{an elementary neuron. As a proof-of-concept, we carry out molecular simulations of two nanofluidic slits reproducing the Hodgkin-Huxley model, and observe spontaneous emission of voltage spikes characteristic of neuromorphic activity}.  
		}
		
	\end{sciabstract}
	
	\rev{
		Neurotransmission relies on the 
		\LB{harmonious} 
		transport of \LB{manifold} ionic species across the cellular membrane \cite{hille_ionic_1978,gerstner_spiking_2002}. In particular, the non-linear, history-dependent dynamics of biological ion channels \cite{hodgkin_components_1952} is key to many neuronal processes. While considerable progress in the design of novel nanofluidic devices has been achieved over the past decade \cite{Bocquet2020,garaj_graphene_2010,lee_coherence_2010,celebi_ultimate_2014,feng_single-layer_2016,secchi_massive_2016,tunuguntla_enhanced_2017}, artificial systems still cannot compete with neurons' ionic machinery. This points to the need of inventing and designing artificial iontronic devices with advanced functionalities \cite{Bocquet2020}. Most notably, an ion-based memristor -- short for memory resistor, an electronic device with hysteretic conductance \cite{chua_memristor-missing_1971,strukov_missing_2008} -- could serve as an elementary building block for ion-based neuromorphic systems.
		
		\NK{In the fabrication of nanofluidic systems}, a milestone has very recently been reached, as planar confinement of an electrolyte down to a single molecular layer (see Fig. 1 \textbf{A} and \textbf B) was demonstrated and investigated experimentally \cite{radha_molecular_2016,esfandiar_size_2017,mouterde_molecular_2019}. Extremely confined systems are promising from a functional point of view \cite{faucher_critical_2019}: as they are most sensitive to the discrete nature of ions \cite{kavokine_fluids_2021,sparreboom_transport_2010}, they are prone to exhibit exotic transport properties \cite{mouterde_molecular_2019,kavokine_ionic_2019,marcotte_mechanically_2020}. Yet, modelling such subnanometric systems requires to go beyond the traditional Poisson-Nernst-Planck (PNP) equations \cite{kavokine_fluids_2021}, as the reduced dimensionality \LB{drastically} affects the nature of charge interactions and results in qualitative changes in the conduction dynamics \cite{kavokine_ionic_2019}. In this work, we develop a theoretical and numerical framework to study such two-dimensional systems, and show how they can be used to build novel iontronic devices inspired by biological neurons.
		
	}

	In bulk electrolytes, ionic transport is essentially linear due to the high dielectric constant of water reducing the strength of ionic interactions. Conversely, in one-dimensional geometries (such as in carbon nanotubes), translational degrees of freedom are extremely constrained, limiting correlation times and potential memory effects. \NK{Here}, we focus on an intermediate geometry: a monolayer of ions, molecularly confined in a subnanometric slit, see Fig. 1. In such strong planar confinement, ions experience not only a reduction in translational freedom, but also stronger electrostatic interactions whose nature is intermediate between that of 3D and 2D systems, which we term `2D$^+$' interactions. 
	
		\begin{figure}
		\centering
		\includegraphics[width=1\linewidth]{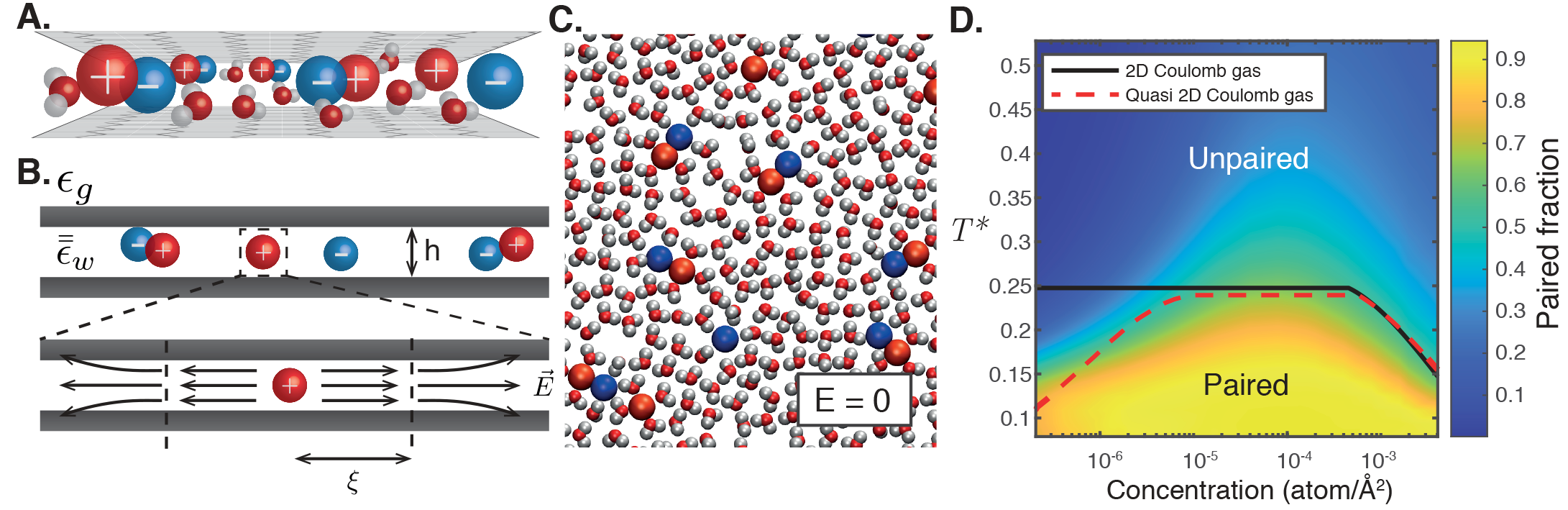}
		\caption{ {\bf Simulation and modelisation of a two-dimensional electrolyte.} {\bf A.} Sketch of an 2D electrolyte. {\bf B.} Quasi-2D Coulomb gas model, with 2D$^+$ Coulomb interactions given by \eqref{eqn:PotQuasi2D}. The electric field is confined in the channel over a length $\xi$ due to dielectric contrast between water and graphite. {\bf C.} Typical configuration of a monolayer electrolyte as observed in all-atom MD simulations. Sodium (in red) and chloride ions (in blue) form tightly bound Bjerrum pairs at room temperature in slits of height $h = 0.7 \, \mathrm{nm}$.  {\bf D.} Phase diagram of a 2D electrolyte, obtained from brownian dynamics simulations. At reduced temperature $T^* = 0.25$, the system undergoes a pairing phase transition, analogous to the Kosterlitz-Thouless transition. The black and red curves correspond to the mean-field prediction for the transition line, for an exactly 2D ($\xi = \infty$) or 2D$^{+}$ (finite $\xi$) Coulomb gas, respectively (see Supplementary Text).}
	\end{figure}
	
	We first investigate the equilibrium properties of 2D electrolytes. Using all-atom MD simulations, we consider aqueous solutions of various salts, such as NaCl, CaCl$_2$ and CaSO$_4$, confined in a narrow slit of tunable spacing $h = 0.7$, $1.0$ or $1.4 \, \si{\nano \meter}$ (amounting to one, two and three water layers, respectively) made of two sheets of graphene or hexagonal boron nitride (hBN). Details of the simulations are provided in the Supplementary Information. 
	While these salts are known to be completely dissociated in bulk solutions, we find that in monolayer confinement, they may form tightly bound Bjerrum pairs \cite{bjerrum_untersuchungen_nodate} (see Fig. 1 \textbf{C}) -- and even triplets in the case of CaCl$_2$. Even in two- or three-layer confinement, all salts except NaCl still associate into pairs.  This pairing was recently predicted for electrolytes confined in one-dimensional carbon nanotubes \cite{kavokine_ionic_2019}, and can similarly be explained in the 2D$^+$ case in terms of the confinement-induced effective interaction between the ions. The water dielectric permittivity $\epsilon_w$ being much larger than that of the confining medium, the electric field lines created by an ion are forced to remain parallel to the channel walls over a typical length $\xi$. An exact computation, reported in the Supplementary Text yields, in the case of a symmetric electrolyte, the corresponding pair-wise interaction potential (see Fig. S3): 
	\begin{equation}
		\beta V_{ij}(r) = -\frac{q_i q_j}{T^*} \log \frac r {r + \xi},
		\label{eqn:PotQuasi2D}
	\end{equation}
	where $\beta$ is inverse temperature, $q_{i,j} = \pm 1$ the charge sign and $1/T^*$ a dimensionless coupling constant. We derive the precise expressions of $V(r)$, $T^*$ and the dielectric length $\xi$ in Supplementary Text (Section 3.1), notably in the experimentally relevant case where the dielectric permittivity of confined water is anisotropic \cite{schlaich_water_2016,fumagalli_anomalously_2018}. In the case of divalent salt in a slit of height $h = 0.7 \, \si{\nano \meter}$ at room temperature, we find typically $T^* = 0.11$ and $\xi = 14 \, \si{\nano \meter}$. The corresponding Bjerrum length  $\ell_{\rm B}$, defined by $\beta V(\ell_{\rm B}) = 1$, is $130 \, \si{\nano \meter}$, which is much larger than the Bjerrum length in bulk water ($\ell_{\rm B} = 0.7~\rm nm$). Therefore, the interaction potential in Eq.~\eqref{eqn:PotQuasi2D} is much stronger than its bulk counterpart $V_{\text{bulk}}(r) = e^2/ 4 \pi \epsilon_0 \epsilon_w r$; this qualitatively explains the confinement-induced ion pairing. The coupling constant $1/T^*$ is also proportional to $Z^2/h$, with $Z$ the valence of ions, explaining why monovalent ions only forms pairs in the thinnest slits.
	
	In order to identify the range of effective temperatures and ionic concentrations in which Bjerrum pairing occurs (Fig. 1 \textbf{D}), we performed implicit solvent brownian dynamics (BD) simulations of a symmetric electrolyte interacting with the derived potential in Eq.~\eqref{eqn:PotQuasi2D}. At short distance ($r\ll \xi$), our interaction potential is logarithmic, and resembles a 2D Coulomb potential. The pairing transition in our monolayer electrolyte is therefore almost described by a 2D Coulomb gas model \cite{levin_electrostatic_2002,minnhagen_two-dimensional_1987}, and is analogous to the Kosterlitz-Thouless (KT) topological phase transition \cite{kosterlitz_ordering_1973}. However, the analogy is not perfect since the interaction potential recovers a bulk $1/r$ behavior at large distances $r \gg \xi$. To account for these observations, we use a mean-field approach inspired by Fuoss's theory of bulk electrolytes \cite{fuoss_conductance_1959}, in which ion pairs are incorporated as a separate species. We are able to determine analytically the pairing transition temperature, both in the ideal 2D and our 2D$^+$ setting (Fig. 1 \textbf{D}), as detailed in the Supplementary Text, Section 3. Our analytical computation reproduces quantitatively the BD simulations, and the ideal 2D description turns out to be valid for all but the lowest salt concentrations. The ion pairing quasi-KT phase transition appears as a specific feature of the monolayer electrolyte. In 1D confinement there can be no phase transition \cite{kavokine_ionic_2019}, and in three dimensions, the ion-ion interactions are usually not strong enough for pairing to occur at room temperature \cite{levin_electrostatic_2002}. \rev{In contrast}, our model predicts a transition temperature $T = 350 \, \si{\kelvin}$ for divalent salts like CaSO$_4$ in slits with $h=1.4 \, \si{\nano \meter}$, or for CaCl$_2$ with $h=0.7 \, \si{\nano \meter}$.
	\par Such strong ionic correlations at equilibrium are at the source of highly non-linear ion transport, illustrated by the current-voltage characteristics obtained from MD simulations (Fig. 2 \textbf{B} and Fig. 2 \textbf{D}). This non-linearity is a signature that conduction proceeds through the breaking of ion pairs. Consistently, a non-linear response is observed for CaSO$_4$, which is paired at equilibrium, unlike NaCl which does not form pairs in the considered conditions (Fig. 2 \textbf{D}).

	This type of ion transport is known as the second Wien effect, as famously pioneered by Onsager \cite{onsager_deviations_1934,kaiser_wien_nodate}. We accordingly develop a theory for the Wien effect in the 2D$^+$ geometry. 
	We consider a chemical equilibrium between pairs and ``free" ions of the form NaCl $\leftrightarrows$ Na$^+$ + Cl$^-$ (see Fig. 2 \textbf{A}). Assuming pairs dissociate with a timescale $\tau_d$ and free ions assemble into pairs with a timescale $\tau_a$, we obtain an evolution equation for the fraction $n_{f}$ of free ions not engaged in a pair:
	\begin{equation}
		\dot n_{f} = \frac{1-n_{f}}{\tau_d} - \frac{n_{f}^2}{\tau_a}.
		\label{eqn:ChemEq}
	\end{equation}
	While computing the dependence of $\tau_d$ with parameters is usually a mathematical challenge, we were able to reduce it to self-similar problem in the particular case of 2D$^+$ confined electrolytes, see Supplementary text, Section 3, yielding:
	\begin{equation}
		\tau_a = \frac{T^*}{4 \pi D \rho},
		\label{eqn:TauA}
	\end{equation}
	\begin{equation}
		\tau_d = \frac{r_0^2}{2 D} \left(\frac{l_E}{r_0}\right)^{1/T^*},
		\label{eqn:IsolatedPair}
	\end{equation}
	where the lengthscale $l_E = k_BT/ZeE$ describes the strength of the external field with respect to thermal fluctuations, $D$ is the diffusion coefficient and $r_0$ is the radius of considered ions. Because this model considers ion pairs as independent, non interacting particles, we refer to it as the isolated pair (IP) model. When the steady state is reached, equation \eqref{eqn:ChemEq} can be solved and the ionic current reads:
	\begin{equation}
		I = N\frac{2ZeD}{k_B T L}E \frac{\tau_a}{2 \tau_d}\left(\sqrt{1 + \frac{2 \tau_d}{\tau_a}}-1\right) \underset{E \to 0}{\propto} E^{a_{IP}(T^*)},
		\label{eqn:ResultingCurrent}
	\end{equation}
	with $L$ being the channel's length and $a_{IP}(T^*) = 1 + 1/2T^*$ the IV curve exponent at low applied voltage. However, this prediction fails to reproduce simulation results, even at very low concentrations, see Fig. 2 \textbf{B} (solid yellow line).
	The all-atom MD simulations provide some hints to understand this discrepancy. As shown on Fig. 2 \textbf{C},
	under an external electric field, ion pairs do not actually break but instead rearrange into gigantic clusters with chemistry-specific size and topology (see Fig. S1 and Supplementary Text, Section 2). These clusters, which we term ``Bjerrum polyelectrolytes", lead to a radically new phenomenology.
	
	\begin{figure}
		\centering
		\includegraphics[width=1\linewidth]{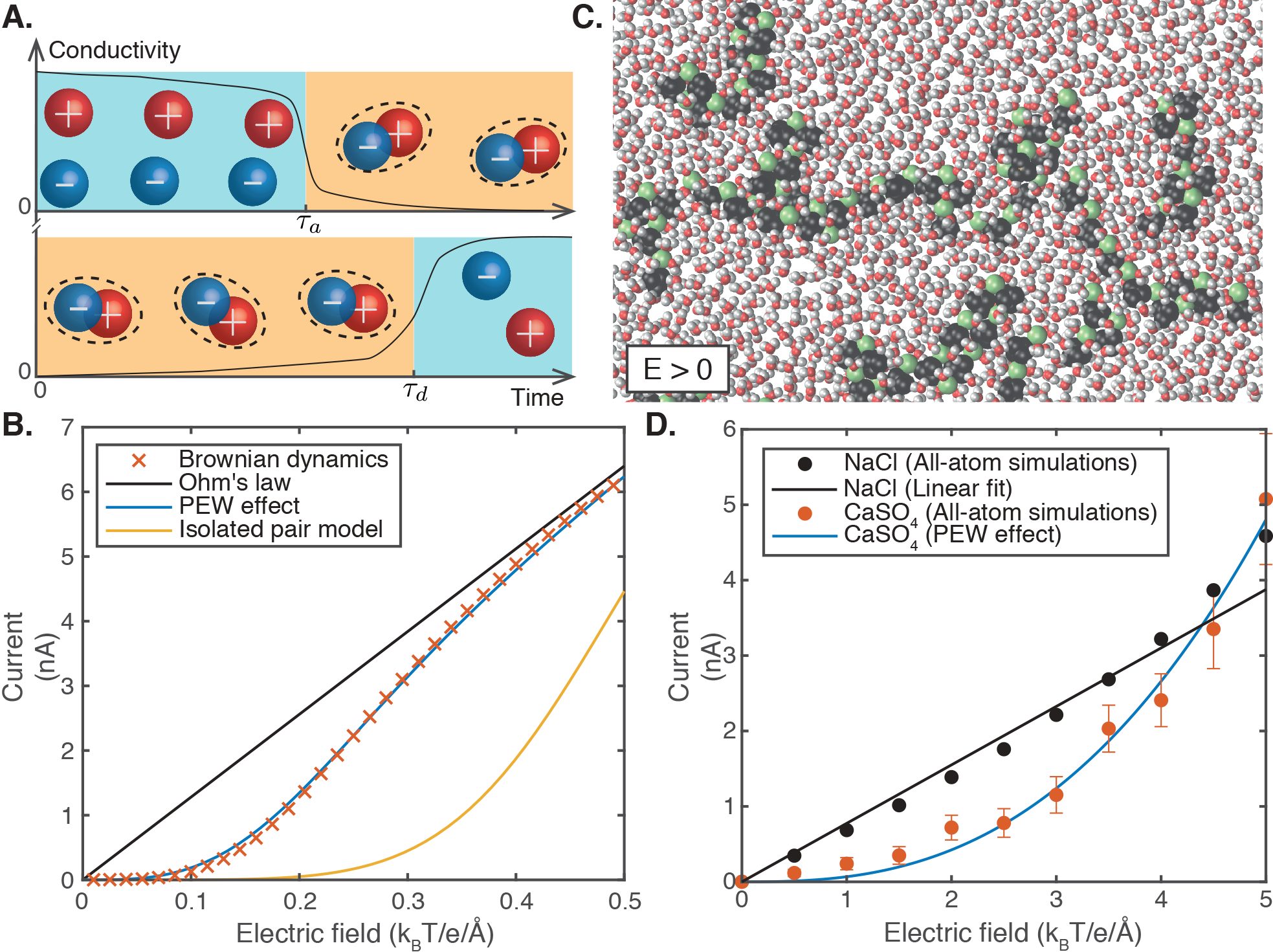}
		\caption{{\bf Analytical theory of the polyelectrolytic Wien (PEW) effect.} {\bf A.} Schematic representation of Onsager's Wien effect for an isolated ion pair: free ions have a lifetime of order $\tau_a$ during which they contribute to conduction, before assembling into pairs (upper panel). Pairs typically break up after a time $\tau_d$, and do not take part in conduction (lower panel).  {\bf B.} Current-voltage characteristic of a generic divalent salt. Taking into account the formation of polyelectrolytes significantly improves the agreement with simulations (blue line, see \eqref{eqn:PEWEffect} and \eqref{eqn:ResultingCurrentPEW}) compared the isolated pair model (yellow line, see \eqref{eqn:IsolatedPair}), for an ionic concentration $\rho = 10^{-3} \, \si{atom/\nano \meter \squared}$. {\bf C.} Under a constant external field, Bjerrum pairs assemble into complex structures: Bjerrum polyelectrolytes, here with CaSO$_4$. Calcium ion are in green and sulfate ions in black. {\bf D.} Current-voltage characteristic of CaSO$_4$ in a $1 \, \si{\nano \meter}$ slit as obtained from all-atom simulations, compared to the power law $I \propto E^a$ of the PEW effect theory, with the predicted exponent $a=a_{PEW}=2.6$. Sodium chloride does not form pairs in slits of that height, and exhibits a linear response.}
	\end{figure}
	
	\par The formation of macrostructures indicates that ion pairs cannot be treated independently, and thus the Wien effect dynamics is fundamentally different from Onsager's picture.
	Going beyond the IP model essentially amounts to take into account the effect of Debye screening, characterized by the Debye length $\lambda_D=\sqrt{\frac{T^*}{\rho n_{f}}}$. Since under a (low) electric field $E$, the free ions density is $n_f \sim \tau_d^{-1/2}\propto l_E^{-1/2T^*}$, then $\lambda_D  \propto l_E^{1/4T^*}$ sets the size of the screening cloud around a given ion. Yet, two ions are carried away from each other if they are separated by a distance exceeding $l_E/T^* \propto l_E$ in the direction of the field, say $x$. For sufficiently low temperature (or sufficiently large field), $\lambda_D > l_E$ and the ionic atmosphere becomes anisotropic: it extends over $l_E$ in the direction $x$ and over $\lambda_D$ in the perpendicular direction $y$. We show in the Supplementary Text (see Section 3.6) that such an anisotropic screening cloud becomes unstable in the direction $y$ below a critical temperature, explaining the formation of Bjerrum polyelectrolytes. In terms of the dynamics, this amounts to modifying the scaling exponent of the pair dissociation time according to (see Supplementary Text, Section 3.6):
	\begin{equation}
		\tau_{d} = \frac{r_0^2}{2 D} \left(\frac{l_E}{r_0}\right)^{1/2T^*}.
		\label{eqn:PEWEffect}
	\end{equation}
	As per Eq.~\eqref{eqn:ResultingCurrent}, this results in a new exponent for the current versus voltage scaling: 
	\begin{equation}
		I  \underset{E \to 0}{\propto} E^{a_{PEW}(T^*)}, \, \, a_{PEW}(T^*) = 1+ 1/4T^*.
		\label{eqn:ResultingCurrentPEW}
	\end{equation}
	We term this new mechanism the polyelectrolytic Wien (PEW) effect, as conduction actually occurs inside the Bjerrum polyelectrolytes.
	In Fig.2 \textbf{B}, we show that the PEW prediction is in quantitative agreement with BD simulation results, \rev{in constrast to} the isolated pair model. 
	We also compare it to results from all-atom simulations for CaSO$_4$ in Fig.2 \textbf{D} to demonstrate its robustness: the dynamics of the ionic assemblies is independent of the chemical nature of the salt and the confining medium or of simulation details (see Fig. S2). Their formation is a direct consequence of 2D confinement.
	
	\begin{figure}
		\centering
	    \includegraphics[width=1.1\linewidth]{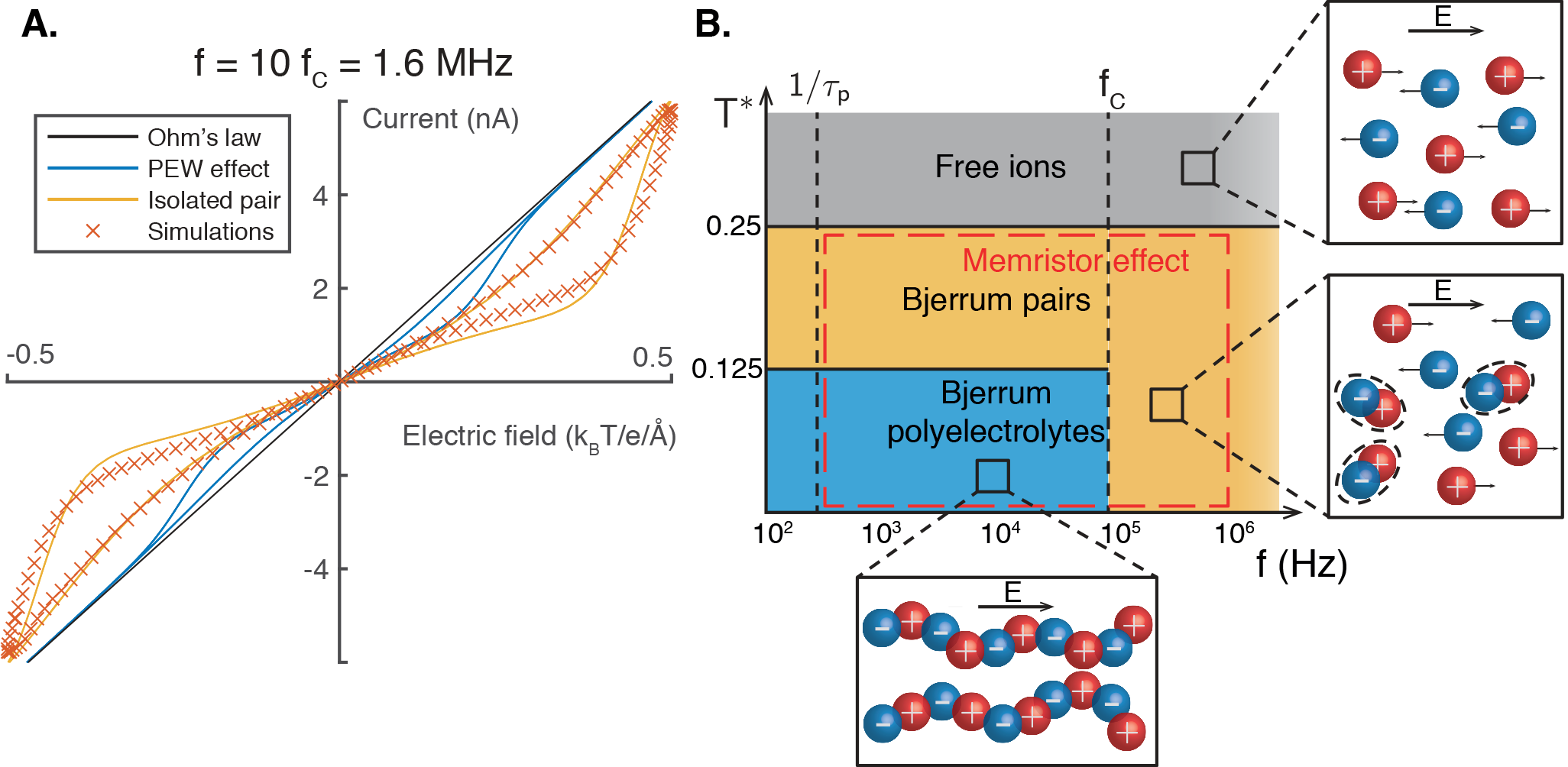}
		\caption{{\bf Time-dependent regime and memristor effect.} {\bf A.} ``Pinched" current-voltage characteristic at high frequency (here $f = 10f_C = 1.6 \, \si{\mega \hertz}$ and $T^* = 0.11$), typical of memristive devices. Here, the IP model (yellow line, equations \eqref{eqn:ChemEq} and \eqref{eqn:IsolatedPair}) provides a better prediction than the PEW effect model (blue line, equations \eqref{eqn:ChemEq} and \eqref{eqn:PEWEffect}), showing that Bjerrum polyelectrolytes are unstable at high frequency. {\bf B.} Temperature-frequency diagram summarizing the existence domains of Bjerrum pairs and polyelectrolytes, for a generic salt with $r_0 = 1\, \si{\angstrom}$, $D=10^{-9} \, \si{\meter \squared \per \second}$ and ionic concentration $\rho \sim 10^{-3} \, \si{atom \per nm\squared}$. Orders of magnitude for other values of parameters are obtained by noticing that the natural scale for frequencies is $D/r_0^2$.}
	\end{figure}
	
	\par The subtle non-linear transport phenomena unveiled above set the stage for the memristor effect. The fraction of ions part of a Bjerrum polyelectrolyte structures is expected to behave as an internal variable, keeping memory of the voltage history. 
	More formally, Eqs. \eqref{eqn:ChemEq} and \eqref{eqn:ResultingCurrent} that can be recast in the more transparent form, introducing the voltage $U$:
	\begin{eqnarray}
		I &=& G(n_{f}) \times U, \label{eqn:Memri1} \\
		\dpart{}{t} n_{f} &=& f(n_{f},U) \label{eqn:Memri2}
	\end{eqnarray}
	The fraction of conducting ions, $n_{f}$, appears here as an internal state variable that depends on the system's history. 
	Furthermore,  from the previous analysis, the relevant timescale governing the system's dynamics -- $\tau_a$ --  is much longer than usual molecular timescales, with typically $\tau_a \sim 1 \, \si{\micro \second}$. 
	This points to potential memory effects in the system. 
	Altogether, Eqs. (\ref{eqn:Memri1}-\ref{eqn:Memri2}) formally defines a voltage-controlled memristor, an electronic device whose resistance depends on its past \cite{chua_memristor-missing_1971,strukov_missing_2008}.
	
	A further proof and hallmark of the memristor behavior is shown in Fig. 3 \textbf{A}, as the I-V characteristic under alternating voltage takes the form of a pinched loop for frequencies larger than a threshold $f_p$.
	The timescale $\tau_p = 1/f_p$ is found to decrease with ionic concentration $\rho$, and we interpret it as a formation timescale of Bjerrum polyelectrolyte. Qualitatively, as conduction occurs through the slow formation of macrostructures, the system cannot adapt to the instantaneous value of the electric \rev{field. Once} the polyelectrolytes \rev{form}, they contribute to conduction even if the field is turned off, before eventually dissolving.
	We compare the BD simulation results with our Wien effect theory to get insight into the underlying physics of the memristor behaviour. We find that below a threshold frequency $f_C$ (here $f_C = 160~\rm kHz$) the simulated I-V curves are well reproduced by the PEW model, while for $f > f_C$, quantitative agreement is observed with the IP model, see Fig. 3 \textbf{B} and Fig. S4. This is consistent with the observation that at high frequencies, polyelectrolytes do not have time to form and conduction proceeds through the breaking of individual pairs. The threshold frequency $f_C = 160~\rm kHz$ is found to be independent of concentration and is therefore related to the dynamics of individual Bjerrum polyelectrolytes. 
	
	While in our simulations the frequency of the alternating field is high due to numerical constraints, our model predicts that it should also be relevant in a realistic experimental setting. We show in the Supplementary Text that a memristor effect could be observed at experimentally relevant frequencies $f \sim 100~\rm Hz$ and voltages $U \sim 0.1~\rm V$ (see Fig. S5). The memory effect is driven by Bjerrum pairs and polyelectrolytes, which form in all salts regardless of their chemical nature. Therefore, the memristor effect, described by equations \eqref{eqn:Memri1} and \eqref{eqn:Memri2}, is a universal property of 2D electrolytes. We also find that optimal conditions correspond to a divalent salt with a slit height close to $1 \, \si{nm}$, with other parameters such as the chemical structure of ions or of the confining material playing little role.
	
	\begin{figure}
		\centering
		\includegraphics[width=1\linewidth]{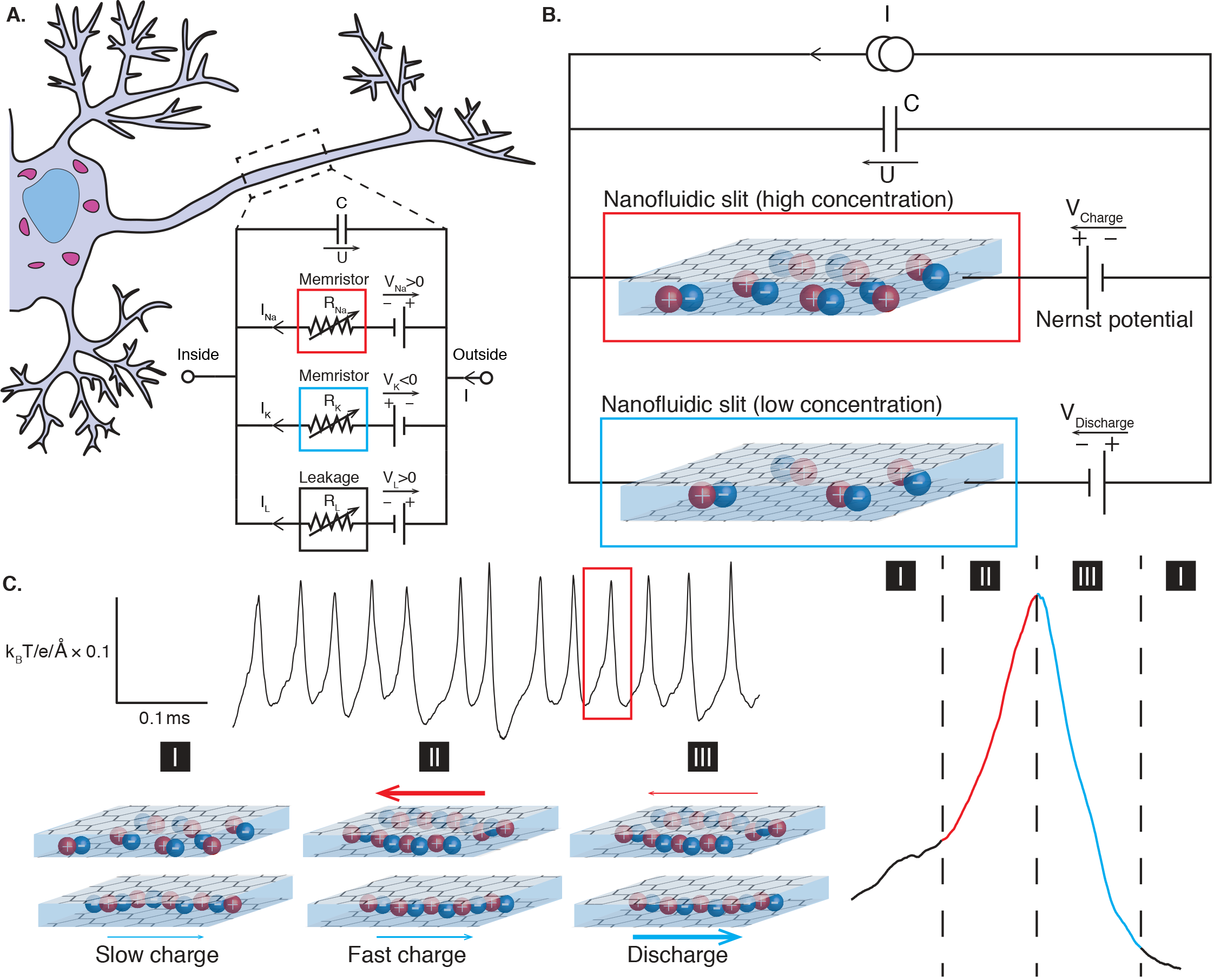}
		\caption{{\bf Building an artificial Hodgkin-Huxley neuron from 2D ionic memristors.} {\bf A.} Electronic representation of the Hodgkin-Huxley model, adapted from \cite{hodgkin_components_1952}. {\bf B.} Prototype ionic machine implemented in BD simulations, exhibiting primitive neuronal behaviour. Two slits with different ionic concentrations are simulated over long timescales ($t \sim 1 \, \si{\milli \second}$). Each slit is connected to a pair of reservoirs , imposing a given Nernst potential on the slit, as in the original computation by Hodgkin and Huxley. {\bf C.} Spontaneous voltage spikes emitted by the prototype ionic machine, and qualitative explanation for the observed spiking effect.}
	\end{figure}
	
	\par Memristors are the electric equivalents of voltage-gated ion channels. As such, they may serve as components of an primitive neuron, as first pointed out by Hodgkin and Huxley \cite{hodgkin_components_1952} (Fig. 4 \textbf{A}). Hence, it is expected that assembling several nanofluidic memristors would allow to mimic neuromorphic behavior. To demonstrate this possibility, we perform parallel BD simulations of two monolayer electrolytes confined in two separate molecular channels, see Fig. 4 \textbf{B}). For both electrolytes, the simulation yields the instantaneous relation between the applied voltage and resulting ionic current. 
	A numeric circuitry is designed around the simulated molecular systems in order to reproduce the  Hodgkin-Huxley model, see Fig. 4 \textbf{A}. It includes a capacitor, and opposite sign potentials applied on each channel that account for the Nernst potentials. 
	\rev{The electronic circuit and the molecular channels are then simulated together, with the capacitor \NK{applying} a voltage drop across the channels.}
	Further details of the simulations are given in the Supplementary Text. We observe spontaneous voltage spikes at a frequency around 10 kHz \rev{(Fig. 4 \textbf B)}. These spikes are the hallmark of neuromorphic behaviour, obtained here from the sole properties of monolayer electrolytes. 
	
	The qualitative mechanism behind the generation of the spikes can be understood as follows. Since the two electrolytes are subject to (Nernst) potentials of opposite signs, they conduct current in opposite directions, but they are gated by the same voltage $U$. When $U$ is small, only the discharging memristor is conducting, so that the capacitor receives a current $I_C < I$ and slowly charges. 
	When $U$ reaches a threshold value, the charging memristor starts conducting and the capacitor receives $I_C \gg I$, causing the voltage to spike up to $U\sim V_{\rm Charge}$. Then, the discharging memristor takes over: the capacitor receives a strongly negative current, and its voltage is lowered back to 0, at which point the process can start again (see Fig. 4 \textbf{C}). This whole process is analogous to voltage spiking in biological neurons caused by the successive opening and closing of ion channels (sodium and potassium channels in the case of the Hodgkin-Huxley model of the giant squid axon).
	The observed working frequency of a few kilohertz is, considering the set of parameters used in simulations, just above the threshold frequency for the memristor effect, which is therefore crucial for observing the neuromorphic behaviour. 
	
	\par The proper design of an experimentally accessible nanofluidic channels \rev{enables reproduction of} the physical processes that occur in an elementary neuron, capable of emitting voltage spike trains. This result builds on the far-from-equilibrium transport properties of electrolytes in molecularly confined nanochannels, whose conductivity highlights a memristor effect. Our findings are supported by molecular simulations, combined with an extensive theoretical framework for the non-equilibrium transport, which generalises Onsager's description of the Wien effect. These properties build on the 2D$^+$ nature of the such channels, where the magnitude of ionic correlations is intermediate between bulk and 1D systems. This theoretical prototyping is the first step towards an experimental demonstration of the ionic memristor, and exploration of memory phenomena in nanofluidic systems in general. The complex interplay between water, surfaces, and ions at the nanoscale gives rise to larger spatial structures -- here polyelectrolytes -- which entail the emergence of slow dynamics and long memory times. 
	
	\clearpage
	
	\bibliographystyle{Science}

	\noindent{\bf Acknowledgements}\\
	The authors thank H. Yoshida for help and discussions on molecular dynamics. 
	
	\noindent\textbf{Funding:} L.B. acknowledges funding from the EU H2020 Framework Programme/ERC Advanced Grant agreement number 785911-Shadoks and ANR project Neptune. 
	This work has received the support of ``Institut Pierre-Gilles de Gennes'', program ANR-10-IDEX-0001-02 PSL and ANR-10-LABX-31. This work was granted access to the HPC resources of CINES under the allocation A0090710395 made by GENCI.
	
	\noindent\textbf{Authors contributions:} L.B. conceived the project. P.R. carried out the theoretical analysis and the molecular dynamics simulations, with inputs from all authors. All authors discussed the results, co-wrote the article and reviewed the final manuscript.
	
	\noindent\textbf{Competing interests:} The authors declare no competing interests.
	
	\noindent\textbf{Data and materials availability:} All data are in the main text or supplementary materials. The molecular dynamics codes that support the plots within this paper and other findings of this study are archived on Zenodo \cite{robin_zenodo_2021}.\\

	\noindent{\bf Supplementary Materials}\\
	Supplementary Text\\
	Fig. S1--S5\\
	Table S1\\
	References \textit{(32--48)}
	
	\clearpage
	
	\clearpage
	\begin{center}
		\Large \bf Supplementary Material
	\end{center}
	
	\ \\
	Supplementary Text\\
	Fig. S1--S5\\
	Table S1\\
	References \textit{(32--48)}
	\clearpage
	
	\section*{Supplementary Text}

	\tableofcontents{}
	
	\clearpage
	
	\section{Methods -- Molecular dynamics}
	
	We use both all-atom molecular dynamics simulations (where water molecules and graphene are simulated explicitly) and Brownian dynamics simulations (where water molecules and graphene are treated implicitly as continuous media). Both are carried out using the LAMMPS software \rev{(08/11/2017 version)} \cite{plimpton_fast_1995}.\par 
	All-atom dynamics simulations consist \rev{of} two graphene sheets separated by a single or a few layers of water, with periodic boundary conditions. This forms a slit of dimensions $L\times w \times h$, with typical values being $L = w = 20 \, \mathrm{nm}$. Channels of three different heights are tested: $h= 0.7$, $1$ or $1.4 \, \mathrm{nm}$, corresponding to one, two and three water layers between the graphene sheets, respectively. The number $N_{w}$ of water molecules is approximately $4500$ for each water layer. This value was obtained from previous studies \cite{yoshida_dripplons_2018}. The number $N$ of ions of each sign is varied between $N=10$ and $N=500$. Simulations involve various binary salts, such as sodium chloride (NaCl), calcium chloride (CaCl$_2$) and calcium sulfate (CaSO$_4$). Generic salts of formula X$_p$Y$_n$ of various valence and stoechiometries were also considered. All initial configurations were obtained using the Packmol software \cite{martinez_packmol_2009}.
	\par Water molecules are modeled using either the TIP4P potential~\cite{jorgensen_comparison_1983} or the SPC/E potential~\cite{berendsen_missing_1987} and maintained rigid with the SHAKE algorithm \cite{ryckaert_numerical_1977}. Lennard-Jones (LJ) parameters of all considered species are summarized in Table S1, with cross-parameters determined with the Lorentz-Berthelot mixing rules \cite{allen_computer_2017}. Interactions are computed using a spherical $1 \, \mathrm{nm}$ cut-off for LJ potentials, and long-range Coulomb interactions are treated with the particle-particle particle-mesh method \cite{hockney_computer_1988} and a slab correction to deal with the non-periodicity in the $z$ direction \cite{yeh_ewald_1999}. Finally, the integration time step is $1 \, \mathrm{fs}$ and temperature is fixed to $298 \, \mathrm{K}$ using the Nos\'e--Hoover thermostat \cite{frenkel_understanding_2001} with a time constant of $0.1 \, \mathrm{ps}$. Simulations last $1.5 \cdot 10^7$ time steps, corresponding to approximately $15 \, \si{\nano \second}$ of physical time.  
	
	In Brownian dynamics simulations, systems consist in $N$ positive and $N$ negative ions, with typically $N = 4000$ in a two-dimensional box of dimensions $L \times w$ and periodic boundary conditions, with $L = 2 \, \si{\micro \meter}$. The width $w$ of the simulation box was fixed depending on the desired electrolyte concentration, with typical values ranging from $w= 0.2 \, \si{\micro \meter}$ to $w= 200 \, \si{\micro \meter}$. The solvent as well as graphene walls are treated implicitly, and the ion positions at timestep $n + 1$ are determined from the positions at timestep $n$ by solving a Euler-discretised overdamped Langevin equation:
	\begin{equation}
		\vecb{r}_{n+1} = \vecb r_n - \Delta t \frac{eD}{k_B T} \vecb \nabla \Phi + \vecb \eta_n \sqrt{2D \Delta t},
	\end{equation}
	where $\vecb r_n$ is the position of a given ion at time step $n$, $\Phi$ is the electrostatic potential felt by the ion and $\vecb \eta_n$ is a Gaussian random variable of zero mean and unit variance. We use the value $D = 10^{-9} \, \si{\meter \squared \per \second}$ for the diffusion coefficient of ions in water at $298 \, \si{\kelvin}$, in line with experimental measurements of the diffusion coefficient of various ions under monolayer confinement \cite{esfandiar_size_2017}.
	
	The typical value of the time step is $\Delta t = 5 \, \si{\pico \second}$, and is lowered down to $ 5 \, \si{\femto \second}$ when fast dynamics are considered. The electrostatic potential is determined by taking into account the contribution $-Ex$ from the external electric field $\vecb E = E \, \unitary x$, as well as pairwise interactions. We assume that ions interact with a pairwise `2D$^+$' quasi-logarithmic potential given by equation (1) of the main text. In typical simulations, we use $T^* = 0.11$ and $\xi = 14 \, \si{\nano \meter}$, which corresponds to divalent ions in a slit of height $h = 7 \, \si{\angstrom}$. Quasi-coulombic, long-range interactions are regularized using both a short- and a long-distance cut-off of $1 \, \si{\angstrom}$ and $20 \, \si{\nano \meter}$, respectively. Simulations last typically $2 \cdot 10^8$ time steps, corresponding to $1 \, \si{\milli \second}$ of physical time. \par
	For both types of simulations, ionic current is computed by averaging the velocities of ions across the slit:
	\begin{equation}
		I = \left\langle \frac{e} L \sum_{i} Z_iq_i\frac{x_{n+1}^i - x_n^i}{\Delta t}\right\rangle_\mathrm{dynamics},
	\end{equation}
	where $q_i = \pm 1$ is the charge of ion $i$, $Z_i$ its valence and the sum is over all ions of both signs. Simulation results were visualized using VMD \cite{humphrey_vmd_1996}.
	
	\section{Pairs, polyelectrolytes and ionic strings}
	\label{sec:IonicAssemblies}
	
	\subsection{Effect of the nature of the salt}
	All-atom MD simulations allow us to study the system's configuration qualitatively. Fig. S1 shows simulation snapshots for all studied chemical species, both at thermal equilibrium (no external field) and out of equilibrium (in presence of an electric field $E \sim 1 \, k_B T\si{/ \angstrom}$). These snapshots correpond to monolayer confinement ($h=0.7 \, \si{\nano \meter}$), except for simulations with sulfate ions, which are too big to enter these slits, and are instead studied in bilayer confinement ($h = 1 \, \si{\nano \meter}$). Sodium chloride forms pairs, or very short clusters (3 to 4 ions) at thermal equilibrium, and assembles into short but tightly bound chains (typically 10 ions or less) under a non-zero electric field. Calcium sulfate mainly forms round clusters (typically 4 to 8 ions) instead of pairs at equilibrium, and very large, branched assemblies for $E\neq 0$. The same is true for a generic divalent salt X$^{2+}$,Y$^{2-}$, except the clusters are linear even at equilibrium, and they almost never branch. 
	
	Calcium chloride, however, behaves slightly differently. It can form either pairs or triplets at equilibrium, or sometimes some short linear assemblies. All these structures are usually not tightly bound, as they include water molecules, solvating calcium ions in a circular arrangement. In the case of triplets, there seem to be different possible value of the dihedral angle formed by the three ions, depending on the number of water molecules surrounding the central cation, or wether that circle is complete or not. Similarly, the Bjerrum polyelectrolytes formed by these triplets have a more complex structure than simple linear chains. We expect that the thermodynamical properties of these triplets and chains to bear some signature of the molecular nature of water; however, the study of these properties is beyond the scope of this work.
	
	Ionic assemblies (pairs and polyelectrolytes) can thus exist in all salts, and they can therefore all exhibit a memristive behaviour, independently of their detailed chemical nature. This proves the robustness of our description, and motivates our focus on Brownian simulations, where chemical details are entirely removed, allowing us to simulate a ``generic'' salt over much longer timescales. However, we believe that these ``details'' are of importance for potential applications of the ionic memristor: by fine-tuning their physical and chemical properties, one could design devices with the same general memristive functionnality, and a wealth of additionnal, application-specific particularities, like biological neurons \cite{gerstner_spiking_2002}.
	
	\subsection{Effect of the nature of confining material}
	
	In Fig. S2, we present simulation snapshots of both NaCl and CaSO$_4$ for two different confining material (graphene and hexagonal boron nitride, or hBN), and two different water models (SPC/E or TIP4P). We observe no qualitative difference between the various combinations of materials and water models, proving that the formation of Bjerrum polyelectrolytes is independent of simulation details, and is not a numerical artefact. 
	
	Graphene and hBN have very similar geometry, but greatly differ in their electronic properties. Graphene is a semiconductor, whose conducting electrons are expected to interact with both water and ions; however, these interactions, which would require a quantum treatment, are well beyong the scope of this work. Hexagonal boron nitride, on the other hand, is an insulator, and therefore is closer to the dielectric confining material considered in our analytical developments. The particular choice of graphene or hBN does not seem to influence the formation of Bjerrum polyelectrolytes, however.
	
	The TIP4P water model is a 4-site model of liquid water (ie., each water molecule consists in 3 atoms and an additional site bearing the negative charge of the oxygen), while SPC/E is a 3-site model. The choice of one over the other was not found to sensibly modify the observed phenomenology.
	
	\section{Coulomb gas model}
	
	Our theoretical analysis of the equilibrium structure of 2D electrolytes is based on the 2D Coulomb gas model \cite{levin_electrostatic_2002,minnhagen_two-dimensional_1987}, which we modify to take into account that graphene slits are not exactly 2D systems. Therefore, we refer to our model as a `2D$^+$' Coulomb gas model. We assume ions to be rigid spheres of radius $r_0$ with a discrete charge $\pm 1$ at their center that can freely move in a 2D plane located at the center of the slit. This approximation is justified as long as the potential created by an ion does not sensibly vary across the slit. For typical values of parameters, we find its relative variation is indeed less than a percent (see equation \eqref{eqn:Series}). Furthermore, ion pairs have a typical size of $4 \, \si{\angstrom}$ in the case of NaCl, exceeding the vertical space available to the electrolyte between the graphene sheets. This greatly constraints their rotational dregrees of freedom, and confirms the validity of our 2D$^+$ approximation
	
	Anions and cations are assumed to have same size, mass, valence and diffusion coefficient $D$. Results for ions with different valence (like CaCl$_2$) can be obtained in a similar way. Lastly, the pairwise interaction potential, derived in next section, is found to be for a $Z:Z$ electrolyte ($Z$ the valence, charge $q_i=\pm Z$ ):
	\begin{equation}
		\beta V_{ij}(r) = -\frac{q_i q_j}{T^*} \log \frac r {r + \xi},
		\label{eqn:SupMatPot}
	\end{equation}
	which strongly resembles the interaction potential of the 2D Coulomb gas, and heavily deviates from Coulomb's law for bulk electrolytes,  which is only recovered (up to a difference in dielectric permittivity, see next section) for $r \gg \xi$. The dielectric confinement length $\xi$ can thus be interpreted as the distance over which electrostatic interactions are effectively 2D. Our computations will sometime consider the case where $\beta V_{ij}(r) = -\frac{q_i q_j}{T^*} \log r$, which we will refer to as the \textit{exact} 2D Coulomb gas (2DCG) model. In many cases -- where $\xi$ happens to be much larger than any other relevant lengthscale -- the 2DCG model provides reasonable predictions, matching the results of molecular dynamics simulations.
	
	In the case of CaCl$_2$, we used the following interaction potential:
	\begin{equation}
		\beta V_{ij}(r) = -\frac{q_i q_j}{T^*} \log \frac r {r + \xi},
	\end{equation}
	with $Z^2 =2$ in the expression of $T^*$ (see Section \ref{sec:DerivationElecPotential}), $q_{i} = +\sqrt{2}$ for calcium ions and $q_i = - 1/\sqrt 2$ for chloride ions. This particular choice allows for a clearer comparison of the various salts studied in function of the value of $T^*$.
	
	\subsection{Coulomb interaction in confinement}
	\label{sec:DerivationElecPotential}
	
	In this section, we derive equation (1) from the main text and relate its parameters $T^*$ and $\xi$ to physical properties of ions and the channel geometry. We also explicit the role of confinement in the properties of electrolytes. In bulk (3D) water, the ion-ion interaction reads:
	\begin{equation}
		\beta V_{ij}^{3D}(r)= \frac{\beta e^2 Z^2 q_iq_j}{4 \pi \epsilon_0 \epsilon_w r},
	\end{equation}
	where $\epsilon_w = 80$ is the relative permittivity of water and $Z$ is the valence of ions. We may define the Bjerrum length $\ell_{Bj} = \beta e^2 Z^2/4 \pi \epsilon_0 \epsilon_w$, which measures the strength of electrostatic interactions with respect to thermal agitation. Analytical study of the 3D Coulomb gas model then shows that ions can form Bjerrum pairs provided their radius is small enough, with $r_0 < r_c = \ell_{Bj}/16$ \cite{levin_electrostatic_2002}. Since $\ell_{Bj} = Z^2 \times 7 \, \si{\angstrom}$ for water at room temperature, this would require $r_0 < 0.4 \, \si{\angstrom}$, and thus the observation of ion pairs in bulk water is impossible for monovalent and divalent ions.
	\par This picture is modified for confined electrolytes. Under strong confinement, the dielectric permittivity of water becomes anisotropic \cite{schlaich_water_2016,fumagalli_anomalously_2018} and its value $\epsilon_\bot$ along the confined direction is greatly reduced, while its value $\epsilon_\parallel$ in other directions remains close to its bulk value $\epsilon_w = 80$. Because the strength of interactions scales with $\sqrt{\epsilon_\bot \epsilon_\parallel}$ at short distances, this anisotropy results in much higher ion-ion correlations, and allows the existence of Bjerrum pairs at room temperature. In what follows, we assume $\epsilon_\bot = 2$ and $\epsilon_\parallel = 80$. Furthermore, the channel is embedded in a confining medium, which we model by a dielectric material of permittivity $\epsilon_m$ typically much lower than that of water. This causes dielectric contrast: the field lines created by an ion remain confined within the channel over a lengthscale $\xi$, which explains the form of \eqref{eqn:SupMatPot}, and, again, reinforces ionic correlations and the stability of ion pairs. In what follows, we assume $\epsilon_m = 2$, but most of our results (and in particular the existence of Bjerrum pairs and polyelectrolytes) can be extended up to $\epsilon_m \sim 10$.
	\par To derive the exact form of the potential, we consider a single point particle of charge $Ze$ located at the center $(r=0,z=0)$ of a slit of height $h$, creating an electrostatic potential $\Phi(r, \theta, z)$. Poisson's equation reads:
	\begin{equation}
		\vecb \nabla (\tensor{\epsilon}\cdot \vecb \nabla \Phi) = - \frac{Ze}{\epsilon_0}\frac{\delta(r) \delta(z)}{2 \pi r},
	\end{equation}
	where $\delta$ is the Dirac distribution and
	\begin{equation}
		\tensor{\epsilon} = 
		\begin{pmatrix}
			\epsilon_\parallel & 0 & 0\\
			0 & \epsilon_\parallel & 0 \\
			0 & 0 & \epsilon_\bot
		\end{pmatrix},
		\label{eqn:dielectricCst}
	\end{equation}
	is the permittivity tensor. For $|z| < h/2$, Poisson's equation becomes:
	\begin{equation}
		\epsilon_\parallel \partial_r^2 \Phi + \frac{\epsilon_\parallel}{r} \partial_r \Phi + \epsilon_\bot \partial_z^2 \Phi = - \frac{Ze}{\epsilon_0}\frac{\delta(r) \delta(z)}{2 \pi r}.
		\label{eqn:Poisson1}
	\end{equation}
	For $|z| > h/2 $, the dielectric constant is isotropic and is equal to $\epsilon_m$, yielding:
	\begin{equation}
		\partial_r^2 \Phi + \frac{1}{r} \partial_r \Phi +  \partial_z^2 \Phi = - \frac{Ze}{\epsilon_0 \epsilon_m}\frac{\delta(r) \delta(z)}{2 \pi r}.
		\label{eqn:Poisson2}
	\end{equation}
	To solve this set of equations, we introduce the Hankel transform $\tilde \Phi$ of the potential:
	\begin{equation}
		\tilde \Phi(k,z) = \int_0^{+\infty} \Phi(r,z) J_0(kr) r \, \di r,
	\end{equation}
	where $J_0$ is the zeroth order Bessel function of the first kind. Equation \eqref{eqn:Poisson1} then reads:
	\begin{equation}
		\partial_z^2 \tilde \Phi - a^2k^2 \tilde \Phi= - \frac{aZe}{\epsilon_0 \sqrt{\epsilon_\bot \epsilon_\parallel}}\frac{ \delta(z)}{2 \pi},
		\label{eqn:Poisson1Transform}
	\end{equation}
	where $a = \sqrt{\epsilon_\parallel/\epsilon_\bot}$. This can be straightfowardly integrated for $|z| < h/2$ into:
	\begin{equation}
		\tilde \Phi(k,z) = \frac{Ze}{4 \pi \epsilon_0 \sqrt{\epsilon_\bot\epsilon_\parallel}} \frac{\exp(-k a|z|)}{k} + A(k) \cosh (k az),
	\end{equation}
	and, likewise for $|z| > h/2$:
	\begin{equation}
		\tilde \Phi(k,z) =  B(k) e^{-k|z|}.
	\end{equation}
	The two integration constants $A$ and $B$ are then determined by imposing the boundary conditions at $z = \pm h/2$. More precisely, the potential must be continuous at $z=\pm h/2$, yielding:
	\begin{equation}
		\tilde \Phi\left(k, z= \frac {h^-}2\right) = \tilde \Phi\left(k, z=\frac {h^+}2\right).
	\end{equation}
	Moreover, upon integration of Poisson's equation over a small rectangular box of height $\delta z \to 0$ and arbitrary width and length, one has:
	\begin{equation}
		\epsilon_\bot \partial_z \tilde{\Phi}\left(k, z=\frac {h^-}2\right) = \epsilon_m \partial_z \tilde{\Phi}\left(k, z=\frac {h^+}2\right).
	\end{equation}
	This yields:
	\begin{equation}
		A = \frac{Ze\exp(-k ah/2)}{4 \pi \epsilon_0 k \sqrt{\epsilon_\bot \epsilon_\parallel}}\frac{\alpha - 1}{\cosh (k ah/2) + \alpha \sinh  (kah/2)},
	\end{equation}
	\begin{equation}
		B = \frac{Ze\exp(-k ah/2)}{4 \pi \epsilon_0 k \sqrt{\epsilon_\bot \epsilon_\parallel}}\Big[1 + \frac{\alpha -1}{1 + \alpha\tanh ( k a h/2)} \Big],
	\end{equation}
	with $\alpha = \frac{\sqrt{\epsilon_\bot \epsilon_\parallel}}{\epsilon_m}$. We can then use Hankel inversion formula:
	\begin{equation}
		\Phi(r,z) = \int_0^{+\infty} \tilde\Phi(k,z) J_0(kr) k \, \di k,
	\end{equation}
	to obtain the exact solution of \eqref{eqn:Poisson1}:
	\begin{equation}
		\Phi(r,z) = \Phi_0(r,z) + \frac{Ze}{4 \pi \epsilon_0 \sqrt{\epsilon_\bot \epsilon_\parallel}}
		\int_0^\infty \di k \, \frac{(\alpha - 1)  \cosh (k az) \exp(-k ah/2)}{\cosh(kah/2) +  \alpha \sinh  (kah/2)}  J_0(rk) ,
		\label{eqn:PotentialExact}
	\end{equation}
	where $\Phi_0$ is the potential created by a point charge in the absence of the confining medium:
	\begin{equation}
		\Phi_{0}(r,z) = \frac{Ze}{4 \pi \epsilon_0 \sqrt{\epsilon_\bot\epsilon_\parallel}}\int_0^\infty  \frac{\exp(-k a|z|)}{k}   k J_0(rk)\, \di k = \frac{Ze }{4 \pi \epsilon_0 \sqrt{\epsilon_\bot\epsilon_\parallel}\sqrt{r^2 + \frac{\epsilon_\parallel}{\epsilon_\bot}z^2}}.
		\label{eqn:CloseRange}
	\end{equation}
	In Brownian dynamic simulations, ions are restricted to the center of the slit, so we only consider the case $z=0$ in what follows. Using the identity:
	\begin{equation}
		\frac{\exp(-kah/2)}{\cosh k a h/2 + \alpha \sinh k a h/2}  =  \frac{2}{\alpha + 1 } \sum_{n = 0}^{+ \infty} \left(\frac{\alpha - 1}{\alpha +1}\right)^n e^{-(n+1)kah},
	\end{equation}
	we finally obtain:
	\begin{equation}
		\Phi(r,z) = \frac{Ze}{4 \pi \epsilon_0 \sqrt{\epsilon_\bot \epsilon_\parallel}} \left[ \Phi_0(r,z) + \sum_{n = 1}^{+ \infty} \left(\frac{\alpha - 1}{\alpha +1}\right)^n \left(\frac{1}{\sqrt{r^2 + a^2 (hn +z)^2}} + \frac{1}{\sqrt{r^2 + a^2 (hn - z)^2}}\right)\right].
		\label{eqn:Series}
	\end{equation}
	This expression contains two corrections compared to the 3D case: the prefactor accounts for the modified permittivity of water, and the second term for dielectric constrast between water and graphite. This expansion, however, is impractical for both simulations and analytical computations. Therefore, we propose a heuristic closed form for $\Phi$. We perform the follwing approximation for $r \gg ah$:
	\begin{equation}
		\Phi(r) \simeq \frac{Ze}{2 \pi \epsilon_0 \sqrt{\epsilon_\bot \epsilon_\parallel}} \sum_{n = 0}^{+ \infty} \frac{\gamma^n}{\sqrt{r^2 + a^2 h^2 n^2}} \simeq \frac{Ze}{4 \pi \epsilon_0 \epsilon_\parallel h} \int_{0}^{\infty} \frac{\gamma^u\, \di u}{\sqrt{r^2/a^2h^2 + u^2}},
	\end{equation}
	with $\gamma = (\alpha -1)/(\alpha +1)$. We obtain:
	\begin{equation}
		\Phi(r) \simeq \frac{Ze}{4 \pi \epsilon_0 \epsilon_\parallel h} \left[-J_0 \left(\frac{r |\log \gamma|}{ah}\right) \log \left(\frac{r |\log \gamma|}{2ah}\right) + \frac 1 2 \pi H_0 \left(\frac{r |\log \gamma|}{ah}\right) - f \left(-\frac{r^2 \log^2 \gamma}{4a^2h^2}\right)\right],
	\end{equation}
	where $H_0$ is the Struve function of zeroth order and $f$ can be expressed in terms of the hypergeometric confluent function ${}_0F_1$ and Euler's gamma function:
	\begin{equation}
		f(x) =\left( \dpart{}{a} \frac{{}_0F_1(a,x)}{\Gamma(a)}\right)_{a=1}.
	\end{equation}
	We can identify the characteristic lengthscale:
	\begin{equation}
		\xi = \frac{a h}{| \log \gamma |} \simeq \frac{\epsilon_\parallel}{2\epsilon_m}h 
		\label{eqn:xidef}
	\end{equation}
	where we recall that $a = \sqrt{\epsilon_\parallel/\epsilon_\bot}$ and $\gamma=(\alpha-1)/(\alpha+1)$ with $\alpha = \frac{\sqrt{\epsilon_\bot \epsilon_\parallel}}{\epsilon_m}$.
	Here one estimates $\xi= 14 \, \si{\nano \meter}$
	for $h = 7 \, \si{\angstrom}$ in the limit of strong dielectric constrast $\epsilon_\parallel \gg \epsilon_m$. The expression of the potential can be further approximated to:
	\begin{equation}
		\Phi(r) \simeq  \frac{-Ze \mathcal K}{2 \pi \epsilon_0 \epsilon_\parallel h} \log \left(\frac{r}{r+\xi}\right),
		\label{eqn:QuasiLog}
	\end{equation}
	which strongly ressembles that of a 2D Coulomb gas of permittivity $\epsilon_\parallel$. The dimensionless constant of order unity $\mathcal{K} \simeq 1.11$ is a geometric factor that depends very weakly on other parameters. All our approximations in theory hold only for $r \gg h$ and strong dielectric constrast, ie. $\gamma$ close to 1. In practice, it provides a good approximation of the interaction potential even for $r$ close to $h$ and $\gamma \simeq 0.7$. Finally, we can write the interaction potential of two ions of charge $q_i$ and $q_j$ at distance $r$:
	\begin{equation}
		\beta V_{ij}(r)=-\frac{q_iq_j}{T^*}\log\left(\frac{r}{r+\xi}\right),
	\end{equation}
	where we introduced the reduced Coulomb temperature:
	\begin{equation}
		T^* = \frac{2 \pi\epsilon_0 \epsilon_\parallel h}{Z^2 e^2 \mathcal K}k_B T.
	\end{equation}
	This concludes our derivation of equation (1) of the main text. The validity of our model can be checked by comparing our 2D$^+$ potential (equation \eqref{eqn:QuasiLog}) to the exact result (equation \eqref{eqn:Series}), see Fig. S3. The agreement is good for all values of $h$ considered in simulations. 
	
	\subsubsection*{Semi-quantitative explanation of the 2D$^+$ regime}
	We now propose a second derivation of the expression of the electrostatic potential, based on orders of magnitude.
	
	Because of dielectric contrast between water and the slit's walls, field lines created by an ion are confined within the slit over a distance $\xi$ around the ion. We thus expect $\Phi \sim -\log r$ for $r<\xi$ (2D regime) and $\Phi \sim 1/r$ for $r > \xi$ (3D regime). Let us consider the cylinder centered on the ion, of radius $\xi$ and height $h$. Flux conservation imposes:
	\begin{equation}
		2 \pi \xi h \epsilon_\parallel E_r \sim 2\pi \xi^2 \epsilon_m E_z,
	\end{equation}
	where $E_r$ and $E_z$ are the components of the electric field in polar coordinates. The axial part $E_z$ corresponds to the potential ``leaking'' through the walls of the slit, while $E_r$ governs the interaction between two ions located on the center plane of the slit. At the transition between the two regimes, the electric field becomes almost isotropic, that is:
	\begin{equation}
		E_r\sim E_z
	\end{equation}
	Therefore, the order of magnitude of the dielectric confinement length is $\xi \sim \frac{\epsilon_\parallel}{\epsilon_m}h$. Lastly, the expression of the electrostatic potential should present a transition from $\Phi \sim - \log r$ to $\Phi \sim 1/r$ for $r \sim \xi$. A natural choice is then:
	\begin{equation}
		\Phi(r) \sim \log \frac{r}{r + \xi}.
	\end{equation}
	
	\subsection{Equilibrium properties}
	
	In this section, we detail under which conditions ionic pairing may occur in 2D electrolytes. The equilibrium properties of the system can be derived using the Debye--Hückel mean-field framework described in \cite{levin_electrostatic_2002} and which we recall here for the sake of completeness. We consider an electrolyte with density $\rho$, with a single ion located at the origin, interacting with all surrounding ions. Assuming ions are rigid spheres of radius $r_0$ that cannot interprenetrate each other, the electrostatic potential created by the test ion is unchanged for $r<r_0$ :
	\begin{equation}
		\Phi(r) \simeq  \frac{-Ze \mathcal K}{2 \pi \epsilon_0 \epsilon_\parallel h} \log \left(\frac{r}{r+\xi}\right) + \psi(Z),
		\label{eqn:PotUnscreened}
	\end{equation}
	with $\psi$ a constant to be determined. For $r>r_0$, the electrostatic potential is given  by the Poisson--Boltzmann (PB) equation, modified to take the special geometry into account:
	\begin{equation}
		\left(\Delta - \frac{1}{r + \xi}\partial_r\right)\Phi = \frac{2Ze\mathcal K \rho}{\epsilon_0\epsilon_\parallel h}\sinh\frac{Ze\Phi}{k_BT}.
	\end{equation}
	The operator on the left-hand side is chosen such that its Green function is given by equation \eqref{eqn:PotUnscreened}. The PB equation must be linearized in order to make computations tractable; however, this underestimates the strength of ion-ion correlations and neglects the possibility of ionic pairing. Bjerrum \cite{bjerrum_untersuchungen_nodate} suggests to introduce pairs as a separate chemical species, decomposing:
	\begin{equation}
		2\rho = \rho_1 + 2\rho_2, 
		\label{eqn:Rho1Rho2}
	\end{equation}
	with $\rho_1$ the concentration in free ions of both signs and $\rho_2$ the concentration in pairs. We get:
	\begin{equation}
		\left(\Delta - \frac{1}{r + \xi}\partial_r\right)\Phi = \frac{Z^2e^2 \mathcal K \rho_1}{\epsilon_0\epsilon_\parallel h k_B T}\Phi = \kappa_D^2 \Phi,
	\end{equation}
	where we introduced the inverse Debye length $\kappa_D$. This leads to:
	\begin{equation}
		\Phi(r>r_0) = \frac{Ze\mathcal KA}{2 \pi \epsilon_0 \epsilon_\parallel h}\left[K_0\left(\kappa_Dr\right) - K_0 \left(\kappa_D\left(r+\xi\right)\right)\right],
	\end{equation}
	where $K_0$ is the modified Bessel function of the second kind. The constants $A$ and $\psi$ are determined using the boundary conditions at $r=r_0$, where $\Phi$ and its gradient must match the unscreened potential \eqref{eqn:PotUnscreened}:
	\begin{equation}
		A = \frac{\xi}{\kappa_D r_0(r_0 + \xi)}\left[K_1\left(\kappa_Dr_0\right) - K_1 \left(\kappa_D\left(r_0+\xi\right)\right)\right]^{-1},
	\end{equation}
	\begin{equation}
		\psi(Z) = \frac{Ze\mathcal K}{2 \pi \epsilon_0 \epsilon_\parallel h}\left[\frac{\xi}{\kappa_Dr_0(r_0 + \xi)}\frac{K_0\left(\kappa_Dr_0\right) - K_0 \left(\kappa_D\left(r_0+\xi\right)\right)}{K_1\left(\kappa_Dr_0\right) - K_1 \left(\kappa_D\left(r_0+\xi\right)\right)}- \log \frac{r_0+\xi}{r_0}\right].
	\end{equation}
	Finally, for a system of size $L\times w$, the electrostatic free energy is obtained through a Debye charging process and reads:
	\begin{equation}
		F^{el} = Lw f^{el} = Lw \rho_1 \int_{0}^{Z} \psi(\lambda) \, \di \lambda.
	\end{equation}
	This allows us to compute the total free energy, which is the sum of this electrostatic term and of entropic terms:
	\begin{equation}
		\beta f = \beta f^{el} + \rho_1\log\frac{\rho_1 \Lambda^2}{2} - \rho_1 + \rho_2\log\frac{\rho_2 \Lambda^2}{\zeta_2} - \rho_2,
		\label{eqn:TotalFreeEnergy}
	\end{equation}
	where $\zeta_2$ is the internal partition function of a pair:
	\begin{equation}
		\zeta_2(R) = \int_{r_0}^{R} 2 \pi r \exp \left[- \frac{1}{T^*}\log \frac{r}{r+\xi}\right].
	\end{equation}
	The large distance cut-off $R$ plays little role in the physics of the system, and can be set to the inflection point of the integral \cite{bjerrum_untersuchungen_nodate,levin_electrostatic_2002}:
	\begin{equation}
		R = \frac{\xi}{T^*}=\frac{\beta Z^2 e^2\mathcal K}{4 \pi \epsilon_0 \epsilon_m} = \ell_{Bj} = 130 \, \si{\nano \meter}.
	\end{equation}
	This lengthscale is called the Bjerrum length. Note that in the limit $T^* \ll 1$, this matches the usual definition of the Bjerrum length $\beta V(\ell_B) = 1$. Its expression is identical to that of bulk water, with the dielectric permittivity of water $\epsilon_w = 80$ replaced by the permittivity of the confining medium $\epsilon_m = 2$. Then, chemical equilibrium between free ions and tightly bound pairs imposes:
	\begin{equation}
		\mu_2 = \mu_+ + \mu_-,
	\end{equation}
	with $\mu_i = \dpart{F}{N_i}$ is the chemical potential of species $i$. This yields:
	\begin{equation}
		\rho_2 = \frac{1}{4}\rho_1^2 \zeta_2 e^{2 \beta \mu^{ex}},
		\label{eqn:Equilibrium}
	\end{equation}
	where we introduced the excess chemical potential defined as:
	\begin{equation}
		\mu^{ex} = \dpart{f^{el}}{\rho_1}.
	\end{equation}
	We can then solve the system comprised of equations \eqref{eqn:Rho1Rho2}, \eqref{eqn:TotalFreeEnergy} and \eqref{eqn:Equilibrium}, by numerically computing the free energy $f$ as a function of ionic concentration $\rho$. At temperature $T^* < T^*_C = 0.25$, $f$ fails to be a convex function of $\rho$, and we define the low and high density branches $\rho_{\rm low}$ and $\rho_{\rm high}$ as the positions of the two inflection points. The curves of $\rho_{\rm low}$ and $\rho_{\rm high}$ as functions of temperature correspond to the spinodal curve. In the 2DCG model $f$ is not convex at low concentration, which corresponds to $\rho_{\rm low}(T^*) = 0$: there is no low density branch. When $T^*$ approaches $T^*_C$, $\rho_{\rm high}(T^*)$ approaches a finite value: in the perfectly 2D case, there is no low density branch, and the high density branch ends at the critical line $T^* = 0.25$ (see Fig. 1 \textbf{D} of the main text, black solid line)
	
	In the 2D$^+$ model, the high density branch is only slightly shifted, but there is a low density branch. However, as we treated ion pairs as an ideal gas, we neglected their contribution to the free energy, and as a result the shape of the low density branch is unphysical. A solution would consist \rev{of} including dipole-ion and dipole-dipole interaction terms, as suggested in \cite{levin_electrostatic_2002}. Since we are only interested in the qualitative shape of the transition line, we simply assume that, at low concentration, the transition happens when there are roughly as many pairs as free ions, ie. $\rho \sim \rho_1$. Using this approximation, we plot the spinodal curve of the 2D$^+$ model in Fig. 1 \textbf{D} (dashed red line). The low and high density branches again do not join, resulting in a critical line at $T^* \simeq 0.25$. 
	
	These results are remindful of the Kosterlitz--Thouless transition of the XY model \cite{kosterlitz_ordering_1973}. The exact value of the critical temperature can be computed in the 2DCG model by considering the $\rho_1 \to 0$ limit:
	\begin{equation}
		\beta f^{el} = - \frac{\rho_1}{T^*}\frac{\log \left[ \kappa_D r_0 K_1 \left(\kappa_D r_0\right)\right]}{(\kappa_Dr_0)^2},
		\label{eqn:FreeELevin}
	\end{equation}
	\begin{equation}
		\beta \mu^{ex} = -\frac{\beta Z^2 e^2\mathcal K}{4 \pi \epsilon_0 \epsilon_\parallel h} \frac{K_0 (\kappa_D r_0)}{\kappa_D r_0 K_1(\kappa_D r_0)} \simeq -\frac{1}{T^*} \Big[ \frac{\gamma}{2} + \frac{\ln \kappa_D r_0 /2}{2} \Big].
		\label{eqn:muex}
	\end{equation}
	We obtain $\rho_2 \propto \rho_1^{2 - 1/2T^*}$. This cannot hold for $T^* < T^*_C = 0.25$: since $\rho_1 < 2 \rho$, we would have $\rho_2 > \rho$ for low enough $\rho$. 
	
	\subsection{Ionic pairing in MD simulations}
	
	As discussed in next section, ionic pairing results in non-linear transport phenomena under an external field. However, a weak enough electric field should not be able to break Bjerrum pairs, and only free ions should contribute to conduction. Therefore, in MD simulations, the paired fraction can be computed from the ionic current at small voltage.
	
	Assuming that pairing captures all relevant ion-ion correlations, free ions essentially behave as independant particles of electrical mobility $\mu_i = Zeq_iD/k_BT$. We obtain:
	\begin{equation}
		I = \left\langle \frac{Z e} L \sum_{\text{free ions}} q_iv_{x}^i\right\rangle = \frac{Z^2 e^2 D w \rho_1}{Lk_B T}\Delta V = \frac{\rho_1}{2 \rho} G_{\text{Ohm}}\Delta V,
		\label{eqn:IVCurveCurrent}
	\end{equation}
	where $G_\text{Ohm}$ is the conductance of an ideal (non interacting) bulk electrolyte of concentration $\rho$. We use this result to define the free and paired fractions in MD simulations:
	\begin{eqnarray}
		n_{f} &=& \frac{G_\text{simulation}}{G_\text{Ohm}}= \frac{\rho_1}{2\rho},\\
		n_{p} &=& 1-n_{f} =  \frac{\rho_2}{\rho}.
	\end{eqnarray}
	This definition of $n_{f}$ is used in Fig. 1 \textbf{D} to obtain the phase diagram of the system from Brownian dynamics. At high temperature, $n_{f} \simeq 1$, and the system is in a purely ohmic regime: there are no pairs and its conductivity matches that of a bulk electrolyte. For $T^* < 0.25$, $n_{f} \simeq 0$ and the system is insulating as there no free ion left within the slit.

	\subsection{Onsager's Wien effect in 2D}
	
	In this section, we derive equations (2) to (5) of the main text and develop an extensive theory of conduction in confined 2D electrolytes far from equilibrium. We follow Onsager's original computation of Wien effect for bulk weak electrolytes \cite{onsager_deviations_1934}, extending it to the 2D case, and then show how it must be modified to take Bjerrum polyelectrolytes into account.
	\par Let us consider a 2D electrolyte (X$^+$, Y$^-$) in an external electric field $\vecb E = E \, \unitary{x}$. We assume ions may pair up according to a chemical equilibrium of the form XY $\leftrightarrows$ X$^+$ + Y$^-$. In what follows, we study the chemical kinetics associated with this equilibrium and we base our model on a generic reaction equation:
	\begin{equation}
		\dot n_{f} = \frac{1-n_{f}}{\tau_d} - \frac{n_{f}^2}{\tau_a},
		\label{eqn:ChemEqSupM}
	\end{equation}
	which corresponds to equation (2) of the main text. The evolution of fraction $n_{f}$ of free ions -- that are not part of a pair and thus can conduct current -- is governed by two timescales: the pair dissociation time $\tau_d$ and the free ions association time $\tau_a$. 
	\par Onsager suggests to derive these two timescales from the shape of the out of equilibrium correlation function $g$. Assuming that a positive ion is fixed at the origin, $g(r,\theta)$ is (up to a normalization factor) the probability density of finding a negative ion at polar position $(r,\theta)$. It is solution of a Fokker--Planck equation:
	\begin{equation}
		\partial_t g = 2D\vecb {\nabla} \cdot ( \vecb {\nabla} g + g \vecb \nabla \left(\beta \Phi\right) ),
		\label{eq:Smoluchowski}
	\end{equation}
	where $\Phi$ is the total electrostatic potential felt by a hypothetical anion located at $(r,\theta)$. One has:
	\begin{equation}
		\beta \Phi =  \frac{1}{T^*}\log\frac{r}{r+\xi} - \frac{r \cos \theta}{l_E}.
		\label{eqn:PhiOnsager}
	\end{equation}
	We recall that $\xi$ is given by \eqref{eqn:xidef}. The first term of the potential corresponds to the unscreened interaction of two ions in confinement: we therefore neglect any influence of the ionic atmosphere or Debye screening. This means the current model only captures the effects of pairs dissociating and ions recombining independently from each others. We therefore call it an \textit{isolated pair model}. The lengthscale $l_E$ introduced in equation \eqref{eqn:PhiOnsager} measures the strength of the external field:
	\begin{equation}
		l_E = \frac{k_B T}{Ze|E|}.
	\end{equation}
	Note that the potential $\Phi$ has its maximum for $r=l_E/T^*$ (plus a correction of order $l_E/\xi \sim 0.01$, which we neglect). This means an ion separated from the central ion by more than $l_E/T^*$ will be carried away by the electric field, breaking the pair. Therefore, $l_E$ is the spatial extension of electrostatic correlations in presence of an external field. The other relevant lengthscale is the Bjerrum length $\ell_{Bj} = \xi/T^*$, which governs the shape of $g$ in the absence of an external field. However, the Wien effect is perceptible in Brownian simulations starting for values of $E$ such that $l_E/\ell_{Bj} \sim 10^{-2} \ll 1$, so we discard any effect caused by a finite dielectric confinement length, and set $\xi = \infty$ in what follows. This amounts to replacing the potential \eqref{eqn:PhiOnsager} by:
	\begin{equation}
		\beta \Phi =  \frac{1}{T^*}\log\frac{r}{r_0} - \frac{r \cos \theta}{l_E} + \text{constant},
	\end{equation}
	where $r_0$ is the ion size. We then assume that the system has reached a steady state, so that $\partial_t = 0$:
	\begin{equation}
		\left[\Delta + \left(\frac{1}{T^*r} - \frac{\cos \theta}{l_E}\right)\partial_r + \frac{\sin \theta}{r l_E} \partial_\theta \right] g = 0.
	\end{equation}
	Since the problem has a single lengthscale $l_E$, we perform the change of variable $\vecb r \to \vecb u = \vecb r/l_E$. We obtain:
	\begin{equation}
		\left[\Delta + \left(\frac{1}{T^*u} - \cos \theta\right)\partial_u + \frac{\sin \theta}{u} \partial_\theta \right]g = 0.
		\label{eqn:SmolReduced}
	\end{equation}
	The system is now entirely determined by a single dimensionless parameter $T^*$, regardless of field strength. This is unique to the 2D case, because of the divergent Bjerrum length. Onsager then suggests to decompose $g$ into two parts:
	\begin{equation}
		g = g_d + g_a,
	\end{equation}
	where $g_d$ and $g_a$ are two solutions of \eqref{eqn:SmolReduced} with the additionnal conditions:
	\begin{equation}
		\int_{0}^{2 \pi }-2D\left[\vecb {\nabla} g_a + g_a \vecb \nabla \left(\beta \Phi\right) \right] \cdot \unitary r \, r \di \theta = -C,
	\end{equation}
	\begin{equation}
		\lim\limits_{r \to \infty}g_a =\rho,
	\end{equation}
	and
	\begin{equation}
		\int_{0}^{2 \pi }-2D\left[\vecb {\nabla} g_d + g_d \vecb \nabla \left(\beta \Phi\right) \right] \cdot \unitary r \, r \di \theta = +C,
	\end{equation}
	\begin{equation}
		\lim\limits_{r \to \infty}g_d =0,
	\end{equation}
	where $C$ is a positive constant independant of $r$. It can be interpreted as a particle flux: $g_a$ describes free ions far from the central ion, with which they can recombine and form a Bjerrum pair. This association process creates a net ionic flux from infinity to the origin; $g_a$ is the solution of \eqref{eqn:SmolReduced} associated with a sink at the origin. Similarly, $g_d$ describes negative ions bound to the central cation and localized near the origin (hence the decay of $g_d$ at infinity), forming a Bjerrum pair. This pair has a certain probability of breaking under the action of the external field, creating an ionic flux from the origin to infinity. These fluxes are of opposite signs and equal amplitude since the system is in steady state.
	\par It is easy to see that in fact:
	\begin{equation}
		g_a = \rho,
	\end{equation}
	and straigthfoward integration yields:
	\begin{equation}
		C = \frac{4 \pi D \rho}{T^*}.
	\end{equation}
	This is a recombination rate, defining the pair association time:
	\begin{equation}
		\tau_a = \frac{T^*}{4 \pi D \rho}.
	\end{equation}
	Under typical settings used in simulations, one has $\tau_a \sim 1 \, \si{\micro \second}$. Computing $g_d$ and the corresponding dissociation timescale $\tau_d$ is, however, a mathematical challenge. Onsager's computation involved a series expansion in terms of families of special functions he invented specifically for this problem; and this solution is applicable if the Smoluchowski equation is spatially separable, which is not the case in 2D \cite{kaiser_wien_nodate}. Instead, we propose an alternative solution based on the self-similarity of $g$, which allows us to derive all the relevant quantities (up to a geometrical factor of order 1) without resorting to in-depth mathematical analysis.
	\par Another trivial solution of \eqref{eqn:SmolReduced} is the Boltzmann distribution:
	\begin{equation}
		g_0(u, \theta) = \exp\left[- \frac 1 {T^*} \ln \frac{ul_E}{r_0} + u \cos \theta \right].
	\end{equation}
	This solution has, however, unphysical behaviour for $u\to \infty$, because the Boltzmann distribution is only relevant at thermal equilibrium. It should, however, bear some physical meaning for $u \ll 1$ because then the effect of the external field is negligible and the system is in quasi-equilibrium. Therefore, we admit that $g_d$ is the unique solution of the following problem:
	\begin{equation}
		\left[\Delta + \left(\frac{1}{T^* u} - \cos \theta\right)\partial_u + \frac{\sin \theta}{u} \partial_\theta\right] g_d= 0,
	\end{equation}
	\begin{equation}
		g_d \sim K_a \left(\frac{l_E}{r_0}\right)^{-1/T^*}u^{-1/T^*} \ \text{for} \  u \to 0,
	\end{equation}
	\begin{equation}
		\lim\limits_{u \to \infty } g_d(u) = 0,
	\end{equation}
	where $K_a$ is chosen such that the particle flux associated with $g_d$ compensates that of $g_a$. Because it is defined through the balance of fluxes, it can be interpreted as an association constant. The uniqueness of the solution is ensured because $g_d$ is known on the whole boundary of the domain. Because $l_E$ now only appears in the boundary condition at $u=0$ as a multiplicative factor, it is easy to see that:
	\begin{equation}
		g_d(u) =  K_a \left(\frac{l_E}{r_0}\right)^{-1/T^*}G(u),
	\end{equation}
	where $G$ is a universal function depending only on $T^*$. The resulting ionic flux reads:
	\begin{equation}
		2DK_a \left(\frac{l_E}{r_0}\right)^{-1/T^*} \mathcal{F}= \frac{4 \pi D \rho}{T^*},
	\end{equation}
	where $\mathcal F$ is the flux associated to the universal function $G$:
	\begin{equation}
		\mathcal F = -\int_{0}^{2 \pi }\left[\vecb {\nabla} G + G \vecb \nabla \left(\beta \Phi_{\vecb u}\right) \right] \cdot \unitary u \, u \di \theta,
	\end{equation}
	the dimensionless potential $\Phi_{\vecb{u}}$ being  defined as:
	\begin{equation}
		\beta \Phi_{\vecb{u}} = \frac{1}{T^*}\log u - u \cos \theta.
	\end{equation}
	The flux $\mathcal F$ has the dimension of an inverse length squared and is independent of $l_E$; it is also independent of $\rho$ the total ionic concentration, since it is the property of a single isolated pair dissociating. Therefore, it must scale with $r_0$, the ionic radius, which is the only relevant lengthscale remaining. We then obtain (up to a geometrical factor of order unity):
	\begin{equation}
		\mathcal F \simeq r_0^{-2}.
	\end{equation}
	We finally get the expression of the association constant $K_a$:
	\begin{equation}
		K_a = 2 \pi \left(\frac{l_E}{r_0}\right)^{1/T^*} \frac{\rho r_0^2}{T^*}=\frac{\tau_d}{\tau_a},
	\end{equation}
	as well as the dissociation time $\tau_d$:
	\begin{equation}
		\tau_d = \frac{r_0^2}{2 D}\left(\frac{l_E}{r_0}\right)^{1/T^*}.
		\label{eqn:WienTauD}
	\end{equation}
	At steady state, the free ion fraction $n_{f}$ can be obtained from \eqref{eqn:ChemEqSupM}:
	\begin{equation}
		n_{f} = \frac{\tau_a}{2\tau_d}\left(\sqrt{1 + \frac{4 \tau_d}{\tau_a}} - 1\right).
		\label{eqn:nfIsolatedPair}
	\end{equation}
	Notably, the above result predicts a power law dependency of $n_{f}$ with applied field at low voltage:
	\begin{equation}
		n_{f} \propto E^{1/2T^*}.
	\end{equation}
	Note that in particular $n_{f}$ vanishes at zero voltage. This shows our model is only valid in the low temperature phase, where all ions are paired up at thermal equilibrium (which is clearly the case in typical simulation settings). Otherwise, Debye screening cannot be neglected, as there remain some free ions even for vanishing electric field. This concludes our solution of the isolated pair model, and we plot its prediction using equations \eqref{eqn:IVCurveCurrent} and \eqref{eqn:nfIsolatedPair} in Fig. 2 \textbf{C} of the main text (yellow line). The agreement with simulations is rather poor. Notably, the onset of conduction happens at much lower voltages than predicted by this model. 
	\par While it provides reasonnable predictions for the bulk Wien effect, the isolated pair model must thus be modified for 2D confined electrolytes. One key element this model fails to account for is the formation of Bjerrum polyelectrolytes at non zero voltages. The existence of ionic strings, as depicted on Fig. 2 \textbf{B}, clearly indicates that the conduction in this system cannot be understood from the individual dynamics of ion pairs. 
	
	\subsection{Role of the ionic atmosphere}
	
	In this section, we discuss the physical meaning of the expression of the dissociation timescale $\tau_d$, so as to extend it to the case where Bjerrum polylectrolytes are relevant. Equation \eqref{eqn:WienTauD} can be recast as an Arrhenius law:
	\begin{equation}
		\tau_d = \tau_{\text{diffusion}}\exp\left[- 2 \beta \Delta F\right] = \frac{r_0^2}{2D}\left(\frac{l_E}{r_0}\right)^{1/T^*},
	\end{equation}
	where $\Delta F$ is the free energy barrier to overcome in order to break a pair. It reads:
	\begin{equation}
		\beta \Delta F \simeq - \frac{\log l_E/r_0}{2T^*}.
	\end{equation}
	This expression is very similar to the free energy cost to create an ionic atmosphere of size $l_E$ around an ion, see \eqref{eqn:FreeELevin}:
	\begin{equation}
		\beta F_{atm}(l_E) = - \frac{1}{T^*}\frac{\log \left[ K_1 \left(r_0/l_E\right) r_0/l_E \right]}{(r_0/l_E)^2} \simeq - \frac{\log l_E/r_0}{2T^*}.
	\end{equation} 
	In other words, the \textit{kinetic} energy barrier to break a pair is equal to the \textit{thermodynamic} energy gap between the paired and the unpaired states, if we admit that the typical size of the ionic atmosphere around a free ion is given by $l_E$ instead of the Debye length $\lambda_D$. 
	\par Indeed, the Debye length reads:
	\begin{equation}
		\lambda_D = \sqrt{\frac{T^*}{2 n_{f} \rho}},
	\end{equation}
	so that for low applied field, we obtain $\lambda_D \propto l_E^{1/4T^*}$. Therefore, $\lambda_D \gg l_E$ provided $T^* < 0.25$, ie. if the system is in the low temperature phase, and for large enough electric field. However, the electrostatic potential \eqref{eqn:PhiOnsager} has a maximum at $r = l_E/T^* \propto l_E$, and thus two ions separated by more than $l_E$ will effectively cease to interact and be carried away by the electric field. This means the ionic atmosphere cannot be larger than $l_E$, and since $\lambda_D \gg l_E$, $l_E$ is indeed its correct lengthscale in our simplified picture.

	\subsection{Anisotropic atmosphere and Bjerrum polyelectrolytes}
	\label{sec:Anisotropic}
	
	The above analysis only holds if the ionic atmosphere is destroyed by the external field. However, we should keep in mind the field only acts along the $x$ axis, while the atmosphere extends in all directions. Hence, correlations along the $y$ axis should remain strong, and the ionic atmosphere becomes anisotropic. It has an ovoid shape, with typical spatial extension $l = \sqrt{\lambda_D l_E}$. Therefore, we also need to compute fluxes along both axes separately. This introduces two quantities, $n_x$ and $n_y$, which are the fractions of ions free to move along the $x$ (resp. $y$) axis. Both follow a reaction equation:
	\begin{equation}
		\dot n_x = \frac{1-n_x}{\tau_{d,x}} - \frac{n_x^2}{\tau_{a,x}},
		\label{eqn:ChemEqNX}
	\end{equation}
	\begin{equation}
		\dot n_y = \frac{1-n_y}{\tau_{d,y}} - \frac{n_y^2}{\tau_{a,y}},
		\label{eqn:ChemEqNY}
	\end{equation}
	where we also introduced distinct association times $\tau_{a,x} = \tau_{a,y} = \tau_a$ and dissociation times $\tau_{d,i}$. Assuming the Arrhenius law derived in last section still holds, the free energy cost to breaking a pair can be decomposed in the following way:
	\begin{equation}
		2 \beta \Delta F (\sqrt{\lambda_{D,y} l_E}) \simeq - \frac{\log l_E/r_0}{2T^*} - \frac{\log \lambda_{D,y}/r_0}{2T^*} = 2\beta \Delta F_x + 2 \beta \Delta F_y,
	\end{equation}
	where $\lambda_{D,y} = \sqrt{T^*/2n_y \rho}$. This allows us to express both dissociation times using Arrhenius equations:
	\begin{equation}
		\tau_{d,x} = \frac{r_0^2}{2D}\left(\frac{l_E}{r_0}\right)^{1/2T^*},
	\end{equation}
	\begin{equation}
		\tau_{d,y} = \frac{r_0^2}{2D}\left(\frac{\lambda_{D,y}}{r_0}\right)^{1/2T^*}.
	\end{equation}
	Chemical equilibrium along the $y$ axis reads:
	\begin{equation}
		\rho(1-n_y)=(\rho n_y)^{2-1/4T^*} \frac{r_0^2}{T^*} \left(\frac{2}{T^*}\right)^{-1/4T^*}.
	\end{equation}
	This equation is the equivalent of \eqref{eqn:Equilibrium}, with temperature divided by 2. Moreover, it can only hold if $T^* > 0.125$. At lower temperatures, the equilibrium is broken and no ion can move freely along the $y$ axis -- remindful of how, at thermal equilibrium, no ion can escape pairing if $T<0.25$. In other words, at low temperature, the whole system collapses into a single (or several) ionic assemblies, which we call Bjerrum  polyelectrolytes. 
	\par Chemical equilibrium along the direction of the applied field reads:
	\begin{equation}
		n_x = \frac{\tau_{a,x}}{2\tau_{d,x}}\left(\sqrt{1 + \left(\frac{2 \tau_{d,x}}{\tau_{a,x}}\right)^2} - 1\right) \underset{E\to 0}{\propto} E^{1/4T^*}.
		\label{eqn:ClusterizedNf}
	\end{equation}
	This last result is similar to \eqref{eqn:nfIsolatedPair}, however the corresponding low field exponent is modified by a factor 2. This shows that Bjerrum polyelectrolytes have a dramatic influence on conduction in confined electrolytes: their formation corresponds to a change of the power law exponent of the current-voltage characteristic, because tearing an ion out of a massive polyelectrolyte is easier than breaking a Bjerrum pair. Our Polyelectrolytic Wien (PEW) effect predicts a greatly increased conductance with respect to Onsager's isolated pair model, as shown on Fig. 2.c of the main text (red solid line), in quantitative agreement with Brownian simulations. 
	
	Note that we mainly focused on ion-ion interactions and did not consider the contribution of pair-ion or pair-pair interactions in this description. While, as stated before, they may be relevant at equilibrium, they can be safely neglected when an electric field is applied. As ions assemble into clusters that entirely govern the conduction dynamics, pairs cease to be relevant to describe the system, and cluster-ion interactions only yield a high-order contribution to the system's energy.
	
	\subsection{Memristor effect}
	
	We now consider the case of an alternating field $E = E_0 \cos 2 \pi f t$ with frequency $f$ ranging from $10 \, \si{\kilo \hertz}$ to $10 \, \si{\mega \hertz}$. An exact theoretical treatment would require to solve the full time-dependent Smoluchowski equation \eqref{eq:Smoluchowski}. This introduces a new lengthscale $l_f = \sqrt{D/f}$ corresponding to a diffusion length over a period of the external field. However, for a typical frequency $f \sim 100 \, \si{\kilo \hertz}$, one has $l_f \sim 100 \, \si{\nano \meter} \gg l_E \sim 1 \, \si{\nano \meter}$. This allows us to neglect terms proportional to $\partial_t g$, of order $(l_E/l_f)^2$, in \eqref{eq:Smoluchowski} and consider quasistatic dynamics. The conclusions of previous sections are hence left unchanged upon replacing $E$ by $E_0 \cos 2 \pi f t$ in the expressions of $\tau_a$ and $\tau_{d,x}$. This quasistatic PEW effect model allows us to predict the shape of the AC IV curve with great accuracy up to a critical frequency $f_c = 160 \, \si{\kilo \hertz}$, see Fig. S4 \textbf{A-C}. At higher frequencies, however, simulations show that Bjerrum polyelectrolytes do not form, and as such the isolated pair model is in better accordance with simulations results compared to the PEW effect model. Moreover, the transition frequency $f_c$ between the two regimes is surprisingly found to be independent of ionic concentration. This frequency must correspond to an intrinsic property of Bjerrum polyelectrolytes, whose formation dynamics are therefore independent of ionic concentration.
	\par The area $\cal A$ of the hysteresis loop measures the strength of the memristor effect, and exhibits a power law dependency with frequency for $f < f_c$. The power law exponent seems to depend logarithmicly on ionic concentration, see Fig. S1 {\bf D}, suggesting an expression of the form:
	\begin{equation}
		\mathcal{A} \sim \exp \left( B \log f \log \rho^{-1}\right).
	\end{equation}
	
	In dilute systems, the memristor effect is weaker, but observable at lower frequencies. This is because electrostatic correlations are smaller, and thus a larger proportion of ions does not engage in pairs or polyelectrolytes, and parcipates linearly to conduction. Note that the overall ionic current is comparable to denser systems, because the total number of particles is kept fixed (to stay in the thermodynamic limit), and only the size of the system varies.
	
	\section{Hodgkin--Huxley neuron model}
	
	In this section, we detail the implementation of the molecular dynamics simulations of the Hodgkin--Huxley neuron model \cite{hodgkin_components_1952} using our ionic memristor. It consists in two graphene slits coupled by an external electronic circuit containing a current generator $I$ and a capacitor $C$, see Fig. 4 of the main text. We refer to these two slits as \textit{discharging} and \textit{charging} memristor due to their respective effect on the capacitor. 
	
	In the description of biological neurons by Hodgkin and Huxley, ion channels (typically sodium and potassium channels) play the role of memristors. Each channel is modelled by a history-dependent resistor in series with a Nernst potential which accounts for the concentration contrast of some ionic species in the reservoirs to which the memristor is connected (the extra- and intracellular mediums), see Fig.4 \textbf{A} of the main text. Hodgkin and Huxley then propose the following electronic model of the ion channels:
	\begin{equation}
		I_k = G_k(n_k)(U-V_k),
	\end{equation}
	\begin{equation}
		\dot n_k = f(n_k,U),
	\end{equation}
	where $I_k$ is the current flowing out of the $k^{\text{th}}$ ion channel, $G_k$ its conductance, $V_k$ its Nernst potential, $U$ the applied voltage and $n_k$ an internal parameter (or array of parameters) describing the activity of the channel. The key point here is that the evolution of $n_k$ only depends on the physical voltage $U$ and not on the Nernst potential, which is of chemical origin. Lastly, since biological channels are ion-specific, their Nernst potential is directly linked to the concentrations of the corresponding ion inside and outside the neuron, $c_k^\text{in}$ and $c_k^\text{out}$:
	\begin{equation}
		V_k = \frac{k_B T}{e} \log \frac{c_k^\text{out}}{c_k^\text{in}}.
	\end{equation}
	Qualitatively, the spiking response observed in the Hodgkin--Huxley model stems from the sign difference in the Nernst potentials of sodium ions (which are more concentrated in the extracellular medium, flow inside the neuron and increase its charge) and potassium ions (which are more concentrated in the intracellular medium, flow outside the neuron and decrease its charge).

	Because our description of 2D electrolytes is formally equivalent to Hodgkin and Huxley's model of ion channels, we can reproduce step by step their neuron model, using two nanofluidic memristors. Illustrating {\it e.g.} the mechanism with CaSO$_4$ as a salt, 
	a ``charging" memristor is connected on the left to a reservoir with more sulfate ions than calcium ions  and on the right to a reservoir containing more calcium than sulfate ions. In both reservoirs, electroneutrality is imposed by some additional electrolyte that cannot enter the graphene slit (because, for example, it is too large). The situation is reversed for the discharging memristor: it is connected on the left to a reservoirs with many calcium ions, and on the right to a reservoir with many sulfate ions.
	\par However in the present simulation setup, we can simplify the setup by imposing {\it de facto} the additional Nernst potential. 
	We cannot indeed afford to simulate the reservoirs explicitly, because we need to perform the simulation over very long timescales (up to a few milliseconds). Instead we implement them through additional Nernst electrochemical potentials $ V_\textrm{Discharge} < 0 $ and $ V_\textrm{Charge} > 0 $ being imposed on the slits.
	
	Our simulation scheme is as follows. First, we impose the capacitor voltage $U(t)$ during a time $\Delta t = 50 \, \si{\nano \second}$ to both slits and compute the corresponding ionic current $I_\textrm{Discharge}(t+\Delta t)$ and $I_\textrm{Charge}(t+\Delta t)$ without taking Nernst potentials into account. We then deduce the channels conductance:
	\begin{equation}
		G_k (t+\Delta t) = \frac{I_k(t+\Delta t)}{U(t)}.
	\end{equation}
	The actual ionic currents are then determined by:
	\begin{equation}
		I_{k,\textrm{Nernst}} = G_k(t+\Delta t) (U(t) - V_k),
	\end{equation}
	\PR{as in the Hodgkin--Huxley model.} This allows us to compute the new value of the voltage:
	\begin{equation}
		U(t+\Delta t) = U(t) + \frac{\Delta t}{C} \left[ I - I_{\textrm{Discharge, Nernst}} - I_{\textrm{Charge, Nernst}} \right].
	\end{equation}
	The plot on Fig. 4  \textbf{C} shows the voltage $U$ as function of time for $I = 0.1 \, \si{\nano \ampere}$, $C = 10^{-4} \, \si{\pico \farad}$ and $V_{\textrm{Charge}} = - V_{\textrm{Discharge}} = 0.2 \, k_B T/\si{\angstrom} $. The graphene slits have the same length $L = 2 \, \si{\micro \meter}$ and different ionic concentrations $\rho_{\textrm{Discharge}} = 10^{-4} \, \textrm{atom}/\si{\nano \meter \squared}$ and $\rho_{\textrm{Charge}} = 10^{-5} \, \textrm{atom}/\si{\nano \meter \squared}$. A more realistic value of $C$ can easily be used instead by reducing the slits length, as only the value of the product $CL^2$ matters, or by considering stacks of identical slits in parallel instead.
	
	\section{Orders of magnitude discussion}
	
	Our results, as discussed in previous sections, are based on molecular dynamics simulations of a graphene slit of size $2 \, \si{\micro \meter}$ and height $ h = 7 \, \si{\angstrom}$, where typical electric fields are of the order of $0.1 \, \si{k_B T \per \angstrom}$. This corresponds to an applied voltage $U \sim 50 \, \si{\volt}$. This is of course unrealistic due to water electrolysis starting at $U = 1.23 \, \si{\volt}$. Similarly, to test the memristor effect we used frequencies in the range $10 \, \si{\kilo \hertz} - 10 \si{\mega \hertz}$ which can hardly be accessed to in experiments due to capacitive effects observed in such nanofluidic systems.
	\par This set of parameters was necessary for simulations, as testing a more reasonable frequency like $f \sim 100 \, \si{\hertz}$ would require to simulate the system for $t \sim 0.1 \, \si{s}$ which is unfeasible due to numerical constraints. Likewise, only the value of the electric field bears relevance to the ionic dynamics. As such, reducing the system size allows to consider much more reasonable voltages. Another possibility would be to use lower ionic concentrations or slightly bigger channels ($ h = 1 \, \si{\nano \meter}$), both of which lowers the electric field needed to observe conduction through Wien effect. 
	\par Our theoretical model, however, allows us to predict that the memristor effect should still be observed in experimental conditions. As a proof of concept, we show in Fig. S5 \textbf{A} the AC IV curve obtained for a sinusoidal voltage of amplitude $U = 0.26 \, \si{V}$ and frequency $f = 160 \, \si{\hertz}$ for a slit of length $L = 100 \, \si{\nano \meter}$ and height $h = 0.7 \, \si{\nano \meter}$ containing an ionic concentration $\rho = 10^{-3} \, \si{atom \per \nano \meter \squared}$. The resulting ionic current is less than $10 \, \si{\pico \ampere}$, which is harder to detect, but the experiment can be implemented using stacks of identical devices in parallel to increase the signal-to-noise ratio significantly. 
	\par Similarly, we can use our model to predict the spiking response of such experimentally available systems. We implement this by solving numerically the Hodgkin--Huxley equations corresponding to the device shown on Fig. 4 \textbf{B} of the main text, with $I = 1 \, \si{\pico \ampere}$, $C = 2 \, \si{\pico \farad}$, $\rho_{\textrm{Charge}} = 20 \rho_{\textrm{Discharge}} = 2 \times 10^{-3} \, \si{atom \per \nano \meter \squared}$, and $T^* = 0.22$ (corresponding to slits of height $h = 1.4 \, \si{\nano \meter}$ and divalent ions). We also use $V_{\textrm{Charge, Discharge}} = \pm 50 \, \si{\milli \volt}$, which corresponds to concentration ratios of 100 between the reservoirs. To compensate for the lower concentration (and hence lower conductivity), the discharging memristor is a stack of 5 identical slits.  The result, showing voltage spike trains with frequency around $7 \, \si{\kilo \hertz}$, is presented on Fig. S5 \textbf{B}.
	
	\clearpage
	
	\begin{figure}
		\centering
		\includegraphics[width=0.7\linewidth]{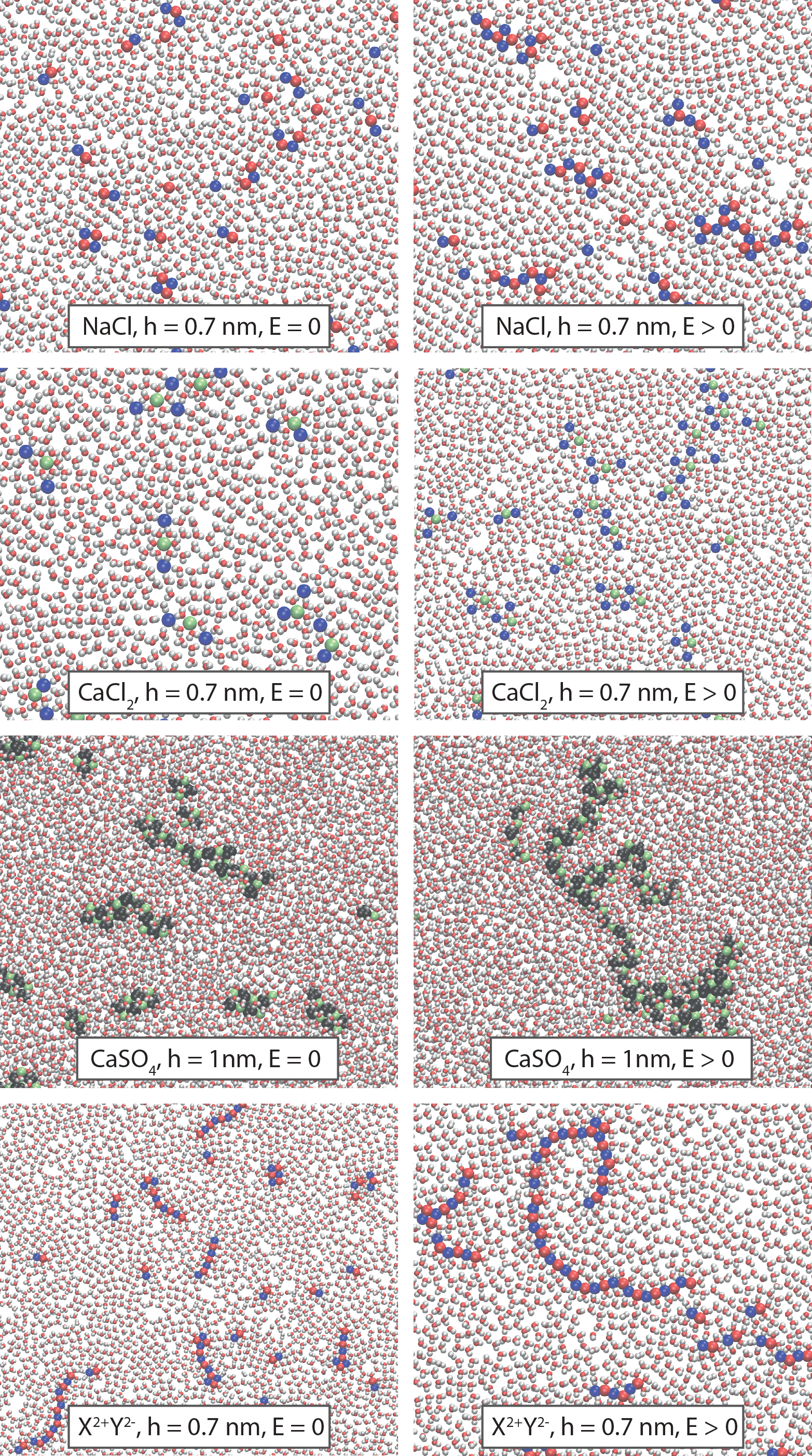}
		\caption*{ Fig. S1 {\bf Simulation snapshots at and out of equilibrium for various salts.} For out of equilibrium snapshots, a constant electric field $E$ of the order of $1 \, k_B T\si{/\angstrom}$ is applied.}
	\end{figure}
	
	\begin{figure}
		\centering
		\includegraphics[width=0.65\linewidth]{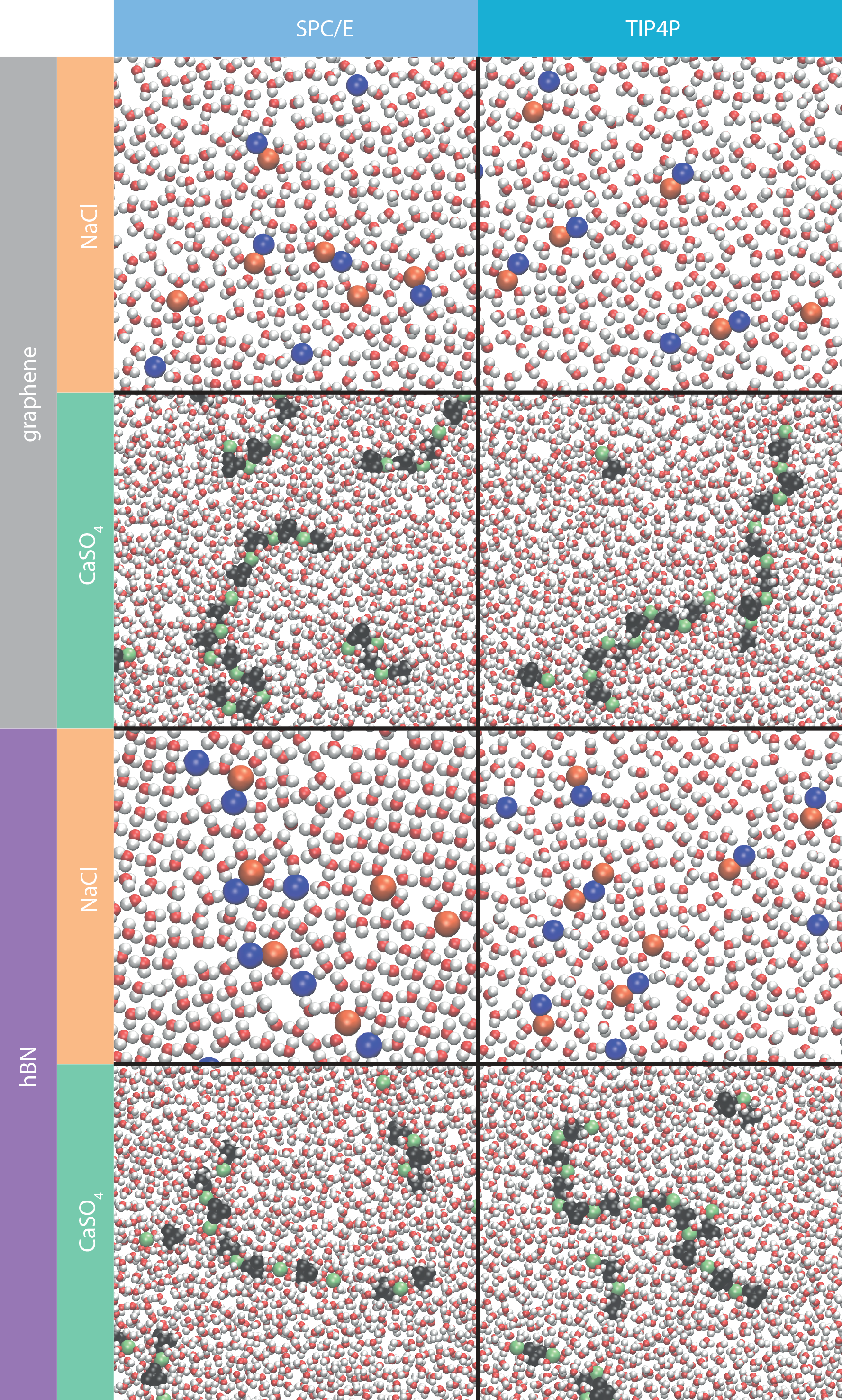}
		\caption*{ Fig. S2 {\bf Comparison of all-atom simulation results for two different confining materials, salts and water models.} We compare the stability of Bjerrum pairs and polyelectrolyte for different simulation settings (graphene or hBN slit, SPC/E or TIP4P water model). Simulations with NaCl correspond to the case of a slit of height $h = 0.7 \, \si{\nano \meter}$, and $h = 1 \, \si{\nano \meter}$ for CaSO$_4$, without any external field. No sensible change is observed in the formation and stability of ionic assemblies, proving the robustness of our description.}
	\end{figure}
	
	\begin{figure}
		\centering
		\includegraphics[width=1\linewidth]{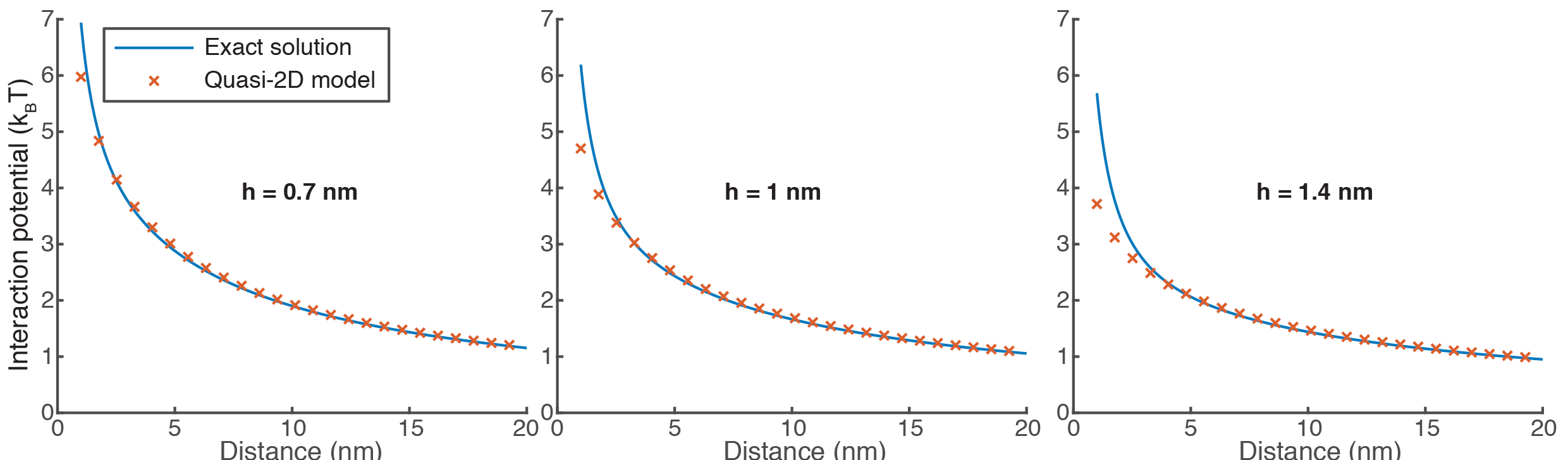}
		\caption*{ Fig. S3 {\bf Comparison between the exact interaction potential and the `2D$^+$' quasi-logarithmic model.} The exact potential is given by equation \eqref{eqn:Series} of Supplementary Text, while the approximate `2D$^+$' model corresponds to equation (1) of the main text. The 2D$^+$ approximation holds up to a few nanometers of confinement, and in particular for the three values of $h$ considered in this work.}
	\end{figure}
	
	\begin{figure}
		\centering
		\includegraphics[width=1.0\linewidth]{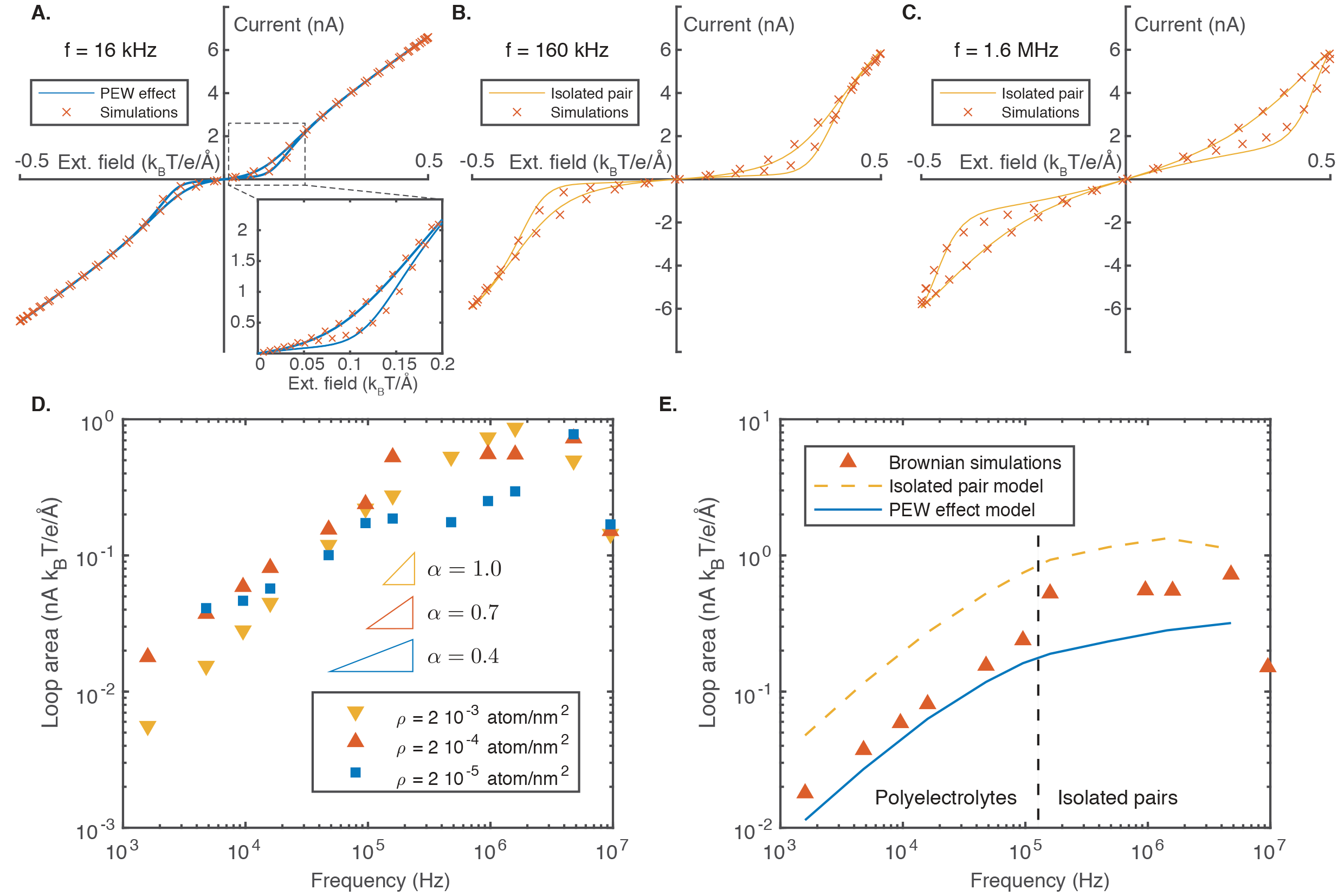}
		\caption*{ Fig. S4 {\bf Characterization of the ionic memristor.} {\bf A.-C.} Low frequency IV curve. The ionic memristor, with ionic concentration $\rho = 10^{-4} \si{atom \per \nano \meter \squared}$ is stimulated with an oscillating field of amplitude $E_0 = 0.5 \, \si{k_B T \per \angstrom}$ and variable frequency. The IV curve exhibits a hysteresis loop for all frequencies in the range $10 \, \si{\kilo \hertz} - 10 \, \si{\mega \hertz}$. At all frequencies ({\bf A.}) below $ f_c = 160 \, \si{\kilo \hertz}$, the shape of the curve can be predicted using a quasistatic version of equation \eqref{eqn:ChemEqNX}. At higher frequencies ({\bf B.}, {\bf C.}) the description of the system in terms of Bjerrum polyelectrolytes fails and the isolated pair model must be used. {\bf D.} Area of the hysteresis loop as function of excitation frequency and ionic concentration. At low frequency, the loop area behaves like a power law of the excitation frequency, which exponents depends logarithmicly on concentration. {\bf E.} Comparison between the two memristor models and Brownian simulations results, showing the shift from polyelectrolytic (PEW effect) to single-pair dynamics (isolated pair model).}
	\end{figure}
	
	\begin{figure}
		\centering
		\includegraphics[width=1.0\linewidth]{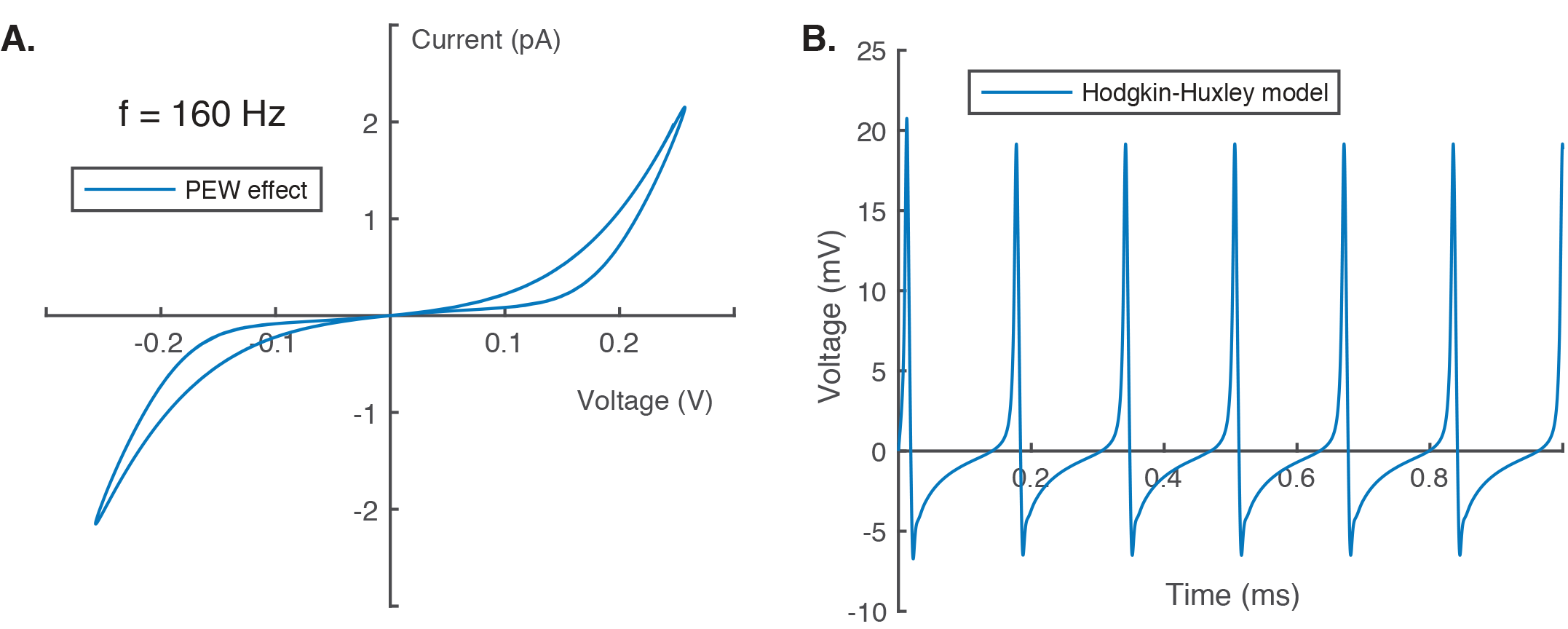}
		\caption*{ Fig. S5 {\bf Theoretical model of experimentally accessible devices.} {\bf A.} Memristor effect in a $100 \, \si{\nano \meter}$ long graphene slit excited by a voltage $U = U_0 \cos 2 \pi f t$ such that $U_0 = 0.26 \, \si{\volt}$ and $f = 160 \, \si{\hertz}$, obtained by solving equation \eqref{eqn:ChemEqNX} numerically. {\bf B.} Voltage spiking in a Hodgkin--Huxley cell (see Fig. 4 \textbf{B} from the main text) made of two $100 \, \si{\nano \meter}$ slits and a capacitance $C = 2 \, \si{\pico \farad}$, stimulated by an input current $I = 1 \, \si{\pico \ampere}$. The two slits are connected to pairs of reservoirs with different cation and anion concentrations, corresponding to the Nernst potentials $V = \pm 50 \, \si{\milli \volt}$. The system is simulated using its equivalent electronic circuit (see Fig. 4 \textbf{A} from the main text).}
	\end{figure}
	
	\clearpage
	
	\begin{table}
		\centering
		\begin{tabular}{@{}lcccc@{}} \toprule
			
			Atom  & Charge $(e)$& $\sigma \, (\si{\angstrom})$ & $\epsilon \, (\si{\kilo \calorie \per \mole})$ & Reference\\ \midrule
			O (H$_2$O) & $-0.8476$ & $3.165$& $0.155$ & \cite{berendsen_interaction_1981} \\
			H (H$_2$O) & $+0.4238$ & -- & -- &  \cite{berendsen_interaction_1981} \\
			C (graphene, SPC/E) & 0 & $3.214$ & $0.0566$ & \cite{werber_materials_2016}  \\
			C (graphene, TIP4P) & 0 & $3.214$ & $0.0566$ & \cite{werber_materials_2016}  \\
			Na$^+$ & $+1$ & $2.35$ & $0.123$ & \cite{smith_computer_1994} \\
			Cl$^-$ & $-1$ & $4.401$ & $0.100$ & \cite{smith_computer_1994} \\
			Ca$^{2+}$ & $+2$ & $2.895$ & $0.100$ & \cite{dang_comment_1995} \\
			S (SO$_4^{2-}$) & $+2$ & $3.55$ & $0.250$ & \cite{cannon_sulfate_1994} \\
			O (SO$_4^{2-}$) & $-1$ & $3.15$ & $0.200$ &  \cite{cannon_sulfate_1994} \\
			B (hBN) & $+0.3$ & $3.402$ & $0.0951$ & \cite{wu_hexagonal_2016}  \\
			N (hBN) & $-0.3$ & $3.490$ & $0.0622$ & \cite{wu_hexagonal_2016}  \\
			X$^{p+}$ & $p=1, 2, 3$ & $2.35$ & $0.123$ & -- \\
			Y$^{n-}$ & $n =1, 2, 3$ & $4.401$ & $0.100$ & -- \\
			\bottomrule
		\end{tabular}
		\caption*{\textbf{Table S1: }Lennard--Jones parameters used in all-atom MD simulations.}
	\end{table}
	
\end{document}